\RequirePackage{lineno}
\documentclass{svjour3}  
\smartqed  %
\usepackage{epsfig}
\usepackage{graphicx}
\usepackage[titletoc,toc,title]{appendix}
\usepackage{lineno}
\usepackage{amsmath}
\usepackage{hyperref}
\usepackage{gensymb}
\usepackage{color}
\usepackage[a4paper,width=130mm,top=25mm, bottom=25mm]{geometry}
\journalname{Journal: Pure and Applied Geophysics}

\newcommand{\myfrac}[2]{\displaystyle \frac{#1}{#2}}

\begin{document}

%\linenumbers

\title{Rock acoustics of diagenesis and cementation}

\subtitle{\normalsize Short title: Seismic physics of rock burial history}

\author{Jos\'e M. Carcione$^{1,2}$ \and Davide Gei$^2$ 
\and Stefano Picotti$^2$ \and Ayman N. Qadrouh$^{3}$ \and  Mamdoh Alajmi$^{3}$
 \and Jing Ba$^{1 (\ast)}$}
\institute{$^1$School of Earth Sciences and Engineering, 
Hohai University, Nanjing, China.  \\
$^2$National Institute of Oceanography and Applied Geophysics Ð OGS, Trieste, Italy. \\
$^3$KACST, PO Box 6086, Riyadh 11442, Saudi Arabia.
\email{jcarcione@inogs.it}
 \\
($\ast$) corresponding author.
\email{jingba@188.com}
}

\date{\today}

\maketitle

\baselineskip 23pt

\begin{abstract}

We simulate the effects of diagenesis, cementation and compaction on the elastic properties of shales and sandstones with four different petro-elastical theories and a basin-evolution model, based on constant heating and sedimentation rates. We consider shales composed of clay minerals, mainly smectite and illite, depending on the burial depth, and the pore space is assumed to be saturated with water at hydrostatic conditions. Diagenesis in shale (smectite/illite transformation here) as a function of depth is described by a 5th-order kinetic equation, based on an Arrhenius reaction  rate. 
On the other hand, quartz cementation in sandstones is based on a model that estimates 
the volume of precipitated quartz cement and the resulting porosity loss from the temperature history, using an equation relating the precipitation rate to temperature. Effective pressure effects (additional compaction) are accounted for by using Athy equation and the Hertz-Mindlin model. The petro-elastic models yield similar seismic velocities, despite the different level of complexity and physics approaches, with increasing density and seismic velocities as a function of depth. The methodology provides a simple procedure to obtain the velocity of shales and sandstones versus temperature and pressure due to the diagenesis-cementation-compaction process. 


\end{abstract}

\section{Introduction}

Diagenesis is the chemical and physical process by which sediments which are buried in the Earth crust are compacted by a process where minerals precipitate from solution due to cementation, grain rearrangement,  with a reduction in porosity (Pytte and Reynolds, 1989; Walderhaug, 1996; Lander and Walderhaug, 1999). Thus, sediments (clay and sand) become rocks by lithification, i.e.,  
the precipitated mineral create bonds between grains, and sand becomes sandstone and clay-rich sediments become shale by this combined diagenesis-cementation process.  
We do not consider the phenomenon by which metamorphic rocks are formed, which typically occurs after that process, beyond 200 $^{\circ}$C, and before melting, roughly 800 $^{\circ}$C, depending on the minerals (Carcione et al., 2018). 
The diagenetic process requires the flow of water in geological time for the minerals to 
precipitate and crystallize. Typical cements include quartz, calcite and clay minerals. 

The acoustics of diagenesis has scarcely been attacked. We can mention Draege et al. (2006), who presented a methodology  
for the estimation of the shale stiffnesses in the
transition zone from mechanical compaction to chemical compaction/cementing. 
The model showed consistent results when comparing
vertical P- and S-wave velocities with log data.
Another type of diagenesis process is the creation of kerogen and bitumen from which
hydrocarbons (oil and gas) are formed by the thermal alteration of these materials. 
A kinetic model can describe this diagenetic process, to model the dissolution-precipitation mechanism (e.g., Luo and Vasseur, 1996;). We do not consider this mechanism in the present work, but it can easily 
be incorporated (Pinna et al., 2011; Carcione et al., 2011; Carcione and Avseth, 2015). 

We consider two dissimilar theories of the diagenesis-cementation process, namely 
smectite/illite conversion based on a kinetic reaction given by the Arrhenius equation (Pytte and Reynolds, 1989), and sandstone compaction and cementation (Walderhaug, 1996; Avseth and Lehocki, 2016). 
The conversion from smectite to illite (clay diagenesis) with increasing depth occurs in all shales (Scotchman, 1987) and can be described by the widely accepted model proposed by Pytte and Reynolds (1989) based on a 5th-order kinetic reaction of the Arrhenius type. The result of the conversion is that the stiffnesses of the minerals composing the shale increase with depth.  
To study quartz cementation in sandstones, one has to estimate cement volumes mainly as a function of temperature. According to Walderhaug (1996), the first stage is precipitation, which occurs at temperatures above 60-80 $^\circ$C, when the cementation starts to be effective. The process depends on the geothermal gradient and on the effective radii of the grains (and surface area), with small quartz grains producing more cement (Walderhaug, 1996, Fig. 4). Moreover, clay minerals or calcite cement covering the grains may reduce the surface area, and inhibit dissolution. 
Roughly, the porosity loss is equal to the volume of precipitated quartz, if the entire cement source is  sourced from outside the sediment volume and a volume of water equal to the cement volume is removed from the sediment. An additional loss is accounted for compaction through the intergranular volume index (IGV) that depends on effective stress
(Lander and Walderhaug, 1999). The model does not consider grain interpenetration due to intergranular pressure solution, which is assumed to be negligible, as supported by experimental evidence (Sippel, 1968; Paxton et al., 2002). However, if required, such an effect can be implemented by using the model of Weyl (1959) or that of Stephenson et al. (1992). 
Figures 1 and 2 show the microstructure of a shale and a sandstone, where the composition and cementation can be appreciated. 
Other illustrative SEM (scanning electron micrographs) images of several shales showing textures of  
montmorillonite (smectite) and illitic shale can be seen in Keller et al. (1986, Figs. 4-22).
SEM images of several sandstones are shown in Walderhaug (2000, Fig 3), with 
high and minor quartz-cement, pores surrounded by almost totally quartz-cemented
areas, and  stylolite containing mica and detrital clay.

We assume a simple  basin-evolution  model  with  a  constant  sedimentation  rate, heating rate  and  geothermal gradient. The diagenesis process starts at a given depth, pressure and temperature, with the  pores  filled  with  liquid water.  Four petro-elastical models yield the seismic velocity of the rocks. The first model (Model 1) is based on an effective mineral 
moduli computed with the Hashin-Shtrikmann average, Krief equation to obtain the dry-rock bulk and shear moduli and Gassmann equations to estimate the wet-rock moduli. Model 2 is based on the self-consistent approximation to obtain the properties of sand/clay mixtures (Gurevich and Carcione, 2000) and partially molten rocks (Carcione et al., 2020). 
Model 3 is a generalization of Gassmann equation to multi-mineral media (Carcione et al., 2005), a theory also used to describe the properties gas-hydrate bearing sediments (Gei and Carcione, 2003). These three models implement Athy equation (Athy, 1930) to account for the pressure effects. 
The fourth approach (Model 4) is a modified version of 
the patchy cement model introduced by Avseth et al. (2016), based on the contact-cement theory (CCT) of Dvorkin and Nur (1996),  assuming cement deposited at grain contacts
(scheme 1 in Mavko et al., 2009, p. 257) (see also Avseth et al., 2010), and the Hertz-Mindlin theory and IGV concept to include pressure effects. Then,  the Voigt-Reuss Hill averages yield the wet-rock moduli, with the self-consistent (SC) model used to obtain the properties of the mineral mixture. 

\section{Temperature-pressure conditions}

Let us assume a rock unit at depth $z$. 
The lithostatic (or confining) pressure for an average sediment density
$\bar \rho$ is $p_c = \bar \rho g z$,
where $g$ is the acceleration of gravity. On the other hand, the hydrostatic 
pore pressure is approximately $p_H = \bar \rho_w g z$, where $\bar \rho_w$ is the density of 
water (values of $\bar \rho$ = 2.5 g/cm$^3$ and $\bar \rho_w$ = 1.04 g/cm$^3$ are assumed here, and 
$g$ = 9.81 m/s$^2$). 

Effective pressure is defined here as 
\begin{equation} \label{pe}
p_e = p_c - p_H , 
\end{equation}
i.e., as differential pressure. 

For a constant sediment burial rate, $S$, and a constant geothermal 
gradient, $G$, the temperature variation of a particular sediment volume is 
\begin{equation} \label{tem}
T = T_0 + G z = T_0 + H t , \ \ \ z = S t , \ \ \ H = G S
\end{equation}
with a surface temperature $T_0$ at time $t$ = 0, where $t$ is deposition time and 
$H$ is the heating rate. 
Typical values of $G$ range from 
20 to 40 $^{\rm o}$C/km, while $S$ may range between 0.02 and 0.5 km/m.y. (m.y. = million
years). 

\section{Diagenetic-compaction processes}

\subsection{Smectite/illite conversion in shales}

To evaluate the amount of smectite/illite ratio forming the shale matrix is important, since this ratio affects the density,  stiffness moduli and wave velocities of the rock. 
Shale mineralogy may include kaolinite, montmorillonite-smectite, illite and chlorite, so
the term smectite/illite as used in this study may be representative for a mixture of clay minerals
(Mondol et al., 2008). 
The smectite/illite composite is  
subject to internal hydration, so its mechanical properties such as the stiffnesses can vary depending on the rock. 

The conversion smectite/illite occurs in all shales with a general release of bound water into the pore space (Scotchman, 1987). Smectite dehydration implies a stiffer matrix due to the presence of more illite and therefore higher velocities. 
The conversion depends on temperature and sedimentation rate. A solution to this problem has been provided by Pytte and Reynolds (1989).  The process is pictorially explained in Figure 3 and mathematical given in Appendix A. 

Future works should consider another effect that could be important, i.e., quartz generation as a by-product of the diagenesis smectite/illite conversion. Thyberg et al. (2009) show that this is another factor to explain the velocity increase, due to micro-crystalline quartz-cementation of the rock frame.

 \subsection{Cementation and compaction in sandstones.}

The physico-chemical process of diagenesis-cementation-compaction in sandstones is shown in Figure 4 and described mathematically in Appendix B by using the theory developed by Walderhaug (1996) for a constant heating rate. 
The model assumes that the quartz grains are spherical and have the same radius.
Surface area decreases as a function of porosity to account for compaction
and cementation. Compaction reduces quartz surface area by increasing grain contact
area and by injecting matrix material
into pore spaces (this last effect neglected here). Cementation can cause further reduction of surface
area when quartz grains are encased by pore-filling cements. 

\section{Petro-elastical models}

We consider four petro-elastical models to obtain the P- and S-wave velocities of the rocks as a function 
of depth, pressure and temperature. In the examples, Model 1, 2 and 3 are applied to shales and all four models to sandstone. Some of the equations, such as Voigt, Reuss and Hashin-Shtrikmann averages, are well-known in the rock-physics community. We refer to Mavko et al. (2009) for details. 

 \subsection{Model 1. Hashin-Shtrikmann-Gassmann model.}

In the first model, smectite and illite (or cement and quartz) are ``mixed" by using the Hashin-Shtrikmann averages to obtain the bulk and shear moduli of the mineral composing the frame, $K_s$ and $\mu_s$ (Appendix C), respectively.  
Then, Krief equation (Krief et al., 1990) yields the dry-rock moduli of the frame with porosity $\phi$, 
\begin{equation} \label{Krief}
K_{m} = K_s ( 1 - \phi)^{A / ( 1 - \phi)} , \ \ \ 
\mu_{m} = \mu_s ( 1 - \phi )^{A / ( 1 - \phi)} , 
\end{equation}
where $A$ is a dimensionless parameter (a value $A$ = 3 is assumed here).

The effect of the pore fluid can be accounted for by using Gassmann equations (e.g., Mavko et al., 2009; Carcione, 2014). The Gassmann bulk and shear moduli are given by  
\begin{equation}\label{Gassmann}
K_G = K_m + \alpha^2 M , \ \ \ \mu = \mu_m, 
\end{equation}
where
\begin{equation} \label{M}
M = \left( \myfrac{\alpha- \phi}{K_s} + \myfrac{\phi}{K_f}\right)^{-1} , 
\end{equation}
\begin{equation} \label{alp}
\alpha = 1 - \myfrac{K_m}{K_s}. 
\end{equation} 
and $K_f$ is the fluid modulus. 

In shales, the reduction in volume due to release of bound water (see Figure 3) is a compaction effect that  is modeled with 
Athy equation, 
\begin{equation} \label{Athy}
\phi = \phi_0 \exp (- \beta p_e)
\end{equation} 
where $\phi_0$ is the initial porosity and $\beta$ is an empirical constant 
(Athy, 1930; Rieke III and Chilingarian, 1974 ) (see also Appendix B) (we consider $\beta$ = 0.01/MPa  in this work). 

 \subsection{Model 2. Self-consistent (SC) scheme.}

In the SC approximation, the elastic moduli of 
an unknown effective medium have to be found implicitly.
The model has been used by Gurevich and Carcione (2000) to obtain the stiffnesses of sand-clay mixtures, where the inclusions are spherical.  
Here, we consider spherical grains (aspect ratio $\gamma$ = 1) and pores of aspect ratio $\gamma < 1$.
In this case, it is $n$ = 3, with smectite, illite and water in the case of shales, and 
quartz, cement and water in the case of sandstones. The proportion of phase 3 ($\phi_3$, the pore space) is given by Athy equation (\ref{Athy}). 

The effective bulk and shear moduli of the composite medium ($K$ and $\mu$), with $n$ phases and proportion $\phi_i$, are obtained as the roots of the following system of equations
\begin{equation} \label{CPA1}
\begin{array}{l}
\sum_{i=1}^n \phi_i (K_i - K) P_i = 0 , \\ \\
\sum_{i=1}^n \phi_i (\mu_i - \mu) Q_i = 0 , \\ \\
\end{array}
\end{equation}
where
\begin{equation} \label{CPA2}
\begin{array}{l}
P_i = \myfrac{K + \frac{4}{3} \mu}{K_i+ \frac{4}{3} \mu} , \ \ \ i = 1, \ldots, n \\ \\
Q_i = \myfrac{\mu + \zeta}{\mu_i+ \zeta}, \\ \\
\zeta = \myfrac{\mu}{6} \cdot \myfrac{9 K + 8 \mu}{K + 2 \mu} 
\end{array}
\end{equation}
for the grains, i.e., $i$ =1, 2 (Mavko et al., 2009, p. 187; Carcione et al., 2020) and 
$P_3 = \frac{1}{3} T_{iijj}$ and $Q_3  = \frac{1}{5} (T_{ijij} - P)$ (for the pores), 
where $T_{iijj}$ and $T_{ijij}$ are given in Appendix A of Berryman (1980) or in page 189 of Mavko et al. (2009). If $\gamma$ = 1, $P_3$ and $Q_3$ are given by equation (\ref{CPA2}). 
A limitation of this theory is that the inclusions are isolated, so that pore pressures are not 
equilibrated and the model computes high-frequency velocities. 

To solve equation (\ref{CPA1}), we use the algorithm developed by Goffe et al. (1994). The Fortran code can be found in: 
\url{https://econwpa.ub.uni-muenchen.de/econ-wp/prog/papers/9406/9406001.txt}.

 \subsection{Model 3. Generalized Gassmann (GG) theory}

We also consider the composite model of Carcione et al. (2005), where the porous medium 
is composed of $n$ = 2 two solids (smectite and illite in shales; quartz and cement in sandstones) and a fluid. If $\phi_i$ is the fraction of the $i$-th solid and $\phi$ is the porosity, it is  $\sum^n \phi_i + \phi$ = 1. The Gassmann modulus is 
\begin{equation} \label{KGgen}
K_G = \sum_{i=1}^n K_{mi} + \left ( \sum_{i=1}^n \alpha_i \right)^n M  ,
\end{equation}
where
\begin{equation} \label{Mgen}
M = \left( \sum_{i=1}^n \frac{\phi_i^\prime}{K_i} + \frac{\phi}{K_f} \right)^{-1} ,
\end{equation}
\begin{equation} \label{effp}
\phi_i^\prime = \alpha_i - \beta_i \phi , \ \ \ \alpha_i =  \beta_i - \frac{K_{mi}}{K_i} , \ \ \ \beta_i = \frac{\phi_i}{1 - \phi} .
\end{equation}
$\beta_i$ is the fraction of solid $i$ per unit volume of total solid. 
Here $K_i$, $i=1, \ldots n$ and $K_f$ are the solid and fluid bulk moduli, respectively, 
and $K_{mi}$, $i=1, \ldots n$ are the frame moduli.

A generalization of Krief's model for a multi-mineral porous medium is used to obtain the frame moduli, 
\begin{equation} \label{Kriefgen}
K_{mi} = (K_s / V_K ) \beta_i K_i ( 1 - \phi )^{A / ( 1 - \phi ) } , \ \ \ \ i = 1, \ldots , n  
\end{equation}
(Carcione et al, 2005),  
$V_K = \sum_{i=1}^n \beta_i K_i$ is the bulk Voigt average, 
and  $K_s$ is given by equation (\ref{HSav}).  
The expression (\ref{Kriefgen}) is such that the composite modulus $K_m = \sum_i^ n K_{mi}$ gives the Hashin-Shtrikmann (HS) average when $\phi$ = 0. 

Similarly,  
\begin{equation} \label{mud1}
\mu_{mi} = (\mu_s / V_\mu ) \beta_i \mu_i ( 1 - \phi )^{A / ( 1 - \phi ) } , \ \ \ \ i = 1, \ldots , n,  
\end{equation}
where $V_\mu = \sum_{i=1}^n \beta_i \mu_i$ is the shear Voigt average,
and  $\mu_s$ is given by equation (\ref{HSav}).  
The dry-rock (and wet-rock) shear modulus of the composite is $\mu_m = \sum_i^ n \mu_{mi}$.

The pressure effects are described by Athy equation (\ref{Athy}). 

 \subsection{Model 4. Patchy cement (PC) model.}

Dvorkin and Nur (1996) developed an elastic model [contact cement theory (CCT)] based on spherical grains (see also Dvorkin et al., 1999).  Let us denote with subscripts 1 and 2 the properties of the grains and cement, respectively. 
The model assumes that initially the sandstone is a random pack
of identical spherical grains with porosity near the critical one ($\phi_0 \approx$  0.36, a critical porosity) and an average number of grain contacts equal to $C$ = 9, with bulk and shear moduli 
\begin{equation} \label{dv1}
K_a = \frac{C}{6} (1-\phi_0) \left( K_2 + \frac{4}{3} \mu_2 \right) S_n  \ \ \ {\rm and} \ \ \
\mu_a = \frac{3}{5} \left[ K_a + \frac{1}{4} {C} (1-\phi_0) \mu_2 S_t  \right], 
\end{equation}
respectively, where ${C}$ is the number of contact per grain and $S_n$ and $S_t$ are 
related to the normal and shear stiffnesses,
respectively, of a cemented two-grain combination. The explicit expressions of these 
quantities are given in Mavko et al. (2009, p. 256) and depend on the Poisson ratio 
of the grains $\nu_1 = (3 K_1-2 \mu_1)/(6 K_1 +2 \mu_1)$, the shear modulus $\mu_1$, and on the ratio of the radius 
of the cement layer to the grain radius:
\begin{equation} \label{dv2}
\alpha = 2 \left[ \frac{\bar \phi_p}{3 {C} (1 - \phi_0)} \right]^{1/4}  ,
\end{equation}
where we consider here that
\begin{equation} \label{dv3}
\bar \phi_p = 
\left\{ 
\begin{array}{ll}
\phi_p, & \ \ \ \phi_p < \phi_{p0} , \\
\phi_{p0}, & \ \ \ \phi_p \ge \phi_{p0} , 
\end{array}
\right.
\end{equation}
where $\phi_p$ is given by equation (\ref{A1}) and $\phi_{p0}$ =0.05 is the maximum amount of cement at the grain contacts (see below and Figure 5). Above this value, the cement is free in the pore space with fraction $\phi_p - \phi_{p0}$. 
Alternatively, for a uniform distribution of the cement on the grain surface, 
\begin{equation} \label{dv21}
\alpha = \sqrt{ \frac{2 \bar \phi_p}{3 (1 - \phi_0)}}  .
\end{equation}
A unified equation for $\alpha$ can be obtained by introducing a new parameter that allows to interpolate between the two cases represented by equations (\ref{dv2}) and (\ref{dv21}) 
(Allo, 2019; Table 1), but as we shall see in the examples, the velocities obtained with these two cases do not differ significantly from a practical viewpoint. 

However, the CCT equation (\ref{dv1}) holds for high porosity, small amounts of cement (at the grain contacts) and does not
consider the effect of pressure. Here, we model pressure effects with a modified version of the patchy cement model of 
Avseth et al. (2016) and include  the compaction effect given by equation (\ref{A7}). The high-porosity limit moduli, $K_b$ and $\mu_b$, are obtained by ``mixing" 
the CCT ($K_a$-$\mu_a$) and Hertz-Mindlin (HM) ($K_u$-$\mu_u$) uncemented moduli (see Appendix D and Figure 6) with the HS upper bound (Appendix C), where the volume fraction of cemented rock to be used in the HS bound is $f = \phi_p/\phi_0$, i.e., $f$ = 0 if there is no cement and 1 if the whole pore-space is filled with cement.

Next, we apply the SC theory to obtain the moduli $K_s$ and $\mu_s$ of a medium where all the pores are filled with cement. There are two phases, grain and cement (as spheres). 
To this purpose, we use equation (\ref{CPA1}) with $n$ = 2, $\phi_1 = 1 - \phi_0$, $\phi_2 = \phi_0$, $K_1 = K_1$,  
$K_2 = K_2$, $\mu_1 = \mu_1$, $\mu_2 = \mu_2$, $K = K_s$ and $\mu = \mu_s$. 

Finally, we interpolate between the effective high-porosity end member given by the HS upper bound and the mineral point (i.e., zero porosity) using the Voigt-Reuss-Hill average,  i.e., an arithmetic average of the Voigt and Reuss moduli, to obtain the dry-rock bulk and shear moduli: 
\begin{equation}\label{kmmm}
K_{m} = \frac{1}{2} ( K_{V} + K_{W}) \ \ \ {\rm and} \ \ \
\mu_m = \frac{1}{2} ( \mu_{V} + \mu_{W}) ,
\end{equation}
where 
%------
\begin{equation}\label{kVkW}
K_{V} = (1 - \phi / \phi_{0}) K_{s} + (\phi/ \phi_{0}) K_{b}  ,  \ \ \ \
\frac{1}{K_{W}} = \frac {1 - \phi / \phi_{0}}{K_{s}} + \frac{\phi/ \phi_{0}}{K_{b}} ,
\end{equation}
%------
\begin{equation}\label{mVmW}
\mu_{V} = (1 - \phi / \phi_{0}) \mu_{s} + (\phi/ \phi_{0}) \mu_{b}   \ \ \ {\rm and} \ \ \
\frac{1}{\mu_{W}} = \frac {1 - \phi / \phi_{0}}{\mu_{s}} + \frac{\phi/ \phi_{0}}{\mu_{b}} .
\end{equation}
%------
The Hill average is close in accuracy to more sophisticated techniques such as self-consistent schemes and are applicable to complex rheologies such as general anisotropy and arbitrary grain topologies (e.g., Man and Huang, 2011) and
anelasticity (Picotti et al., 2018; Qadrouh et al., 2020). 

The wet-rock moduli are obtained with Gassmann equations (\ref{Gassmann}). 

 \section{Seismic velocities}

The P-wave modulus and velocity are 
\begin{equation} \label{Emod}
E = K_G + \frac{4}{3} \mu_m .
\end{equation}
and
\begin{equation} \label{vP}
V_P= \sqrt{\frac{E}{\rho}} ,
\end{equation}
respectively, where $\rho$ is the composite density, given by
\begin{equation} \label{den}
\rho = (1 - \phi) \sum_{i=1}^n \beta_i \rho_i + \phi \rho_f , 
\end{equation}
where $\rho_i$ and $\rho_f$ are the densities of the $i$-th solid phase and fluid, respectively. 

Similarly, the S-wave velocity is 
\begin{equation} \label{vS}
V_S= \sqrt{\frac{\mu}{\rho}} ,
\end{equation}

\section{Examples}

The elastic properties of the minerals are given in Table 1. 
We assume $G$ = 30 $^{\rm o}$C/km, $S$= 0.04 km/m.y, which give 
a heating rate $H$ = 1.2 $^{\rm o}$C/m.y. and we take $T_0$ = 15 $^{\rm o}$C at $z$ = 0.
Figure 7 shows the relation between depth temperature and pressure, corresponding 
to this simple (linear) basin modeling, given by equations (\ref{pe}) and (\ref{tem}). 

 \subsection{Shales}

We consider an initial smectite/illite ratio $r_0$ = 0.99 and $n$ = 5. The kinetic reaction corresponding to smectite/illite conversion assumes 
$E$ = 36 kcal/mol  and $c$ = 1.217 $\times$ 10$^{23}$/ m.y. (Pytte and Reynolds, 1989). 
Figure 8 shows the effect of the geothermal gradient on the illite-smectite fraction, indicating 
that a higher value accelerates the conversion. 
The conversion ratio and density (a) and P- and S- wave velocities (b) of the mineral mixture, as a function of  depth, are shown in Figure 9, where, in this case $G$ = 30 $^{\rm o}$C/km. 
As can be seen, the elastic properties increases with depth due to the higher stiffness and density of illite compared to smectite. 

Next, we consider that at $z$ = 0, the initial porosity is $\phi_0$ = 0.35, and $\beta$ =  0.01/MPa in Athy equation (\ref{Athy}). Then, we compute the porosity, density and wave velocities for Models 1, 2 and 3 as a function of depth, where $\gamma$ = 0.17 in Model 2 (aspect ratio of the pores). Figure 10 shows the porosity and density, which have the same values for the three models, whereas Figure 11 displays the P-wave (a) and S-wave (b) velocities, where the three  models yield similar results in practice. The results of Model 2 differ at lower depths, whose theory is  based on the assumption of idealized geometries of the grains and pores.  If we consider spheres, i.e., $\gamma$ = 1, the P and S-wave velocities predicted by Model 2 are higher by approximately 0.5 km/s than those of  Models 1 and 3. 

To illustrate the effect of the geothermal gradient (and heating rate) on the velocities, we compare the results from two values, $G$ = 20 and 30 $^{\rm o}$C/km ($H$ = 0.8 and 1.2 $^{\rm o}$C/m.y.), keeping the same sedimentation rate. Figure 12 shows the velocities as a function of depth, where the black curves correspond to the higher value of the heating rate (more illite implies higher velocities). The lower velocities for 
20 $^{\rm o}$C/km are due to the higher amount of smectite.

 \subsection{Sandstones}

The dissolution and precipitation of dissolved silica are the source of cement, which
is quantified by equation (\ref{A1}) at geological time. The porosity is given by equation (\ref{A8}), taken into account the pressure effects on compaction. 

We assume the same geothermal gradient, sedimentation rate and heating rate of the first examples, so that Figure 7 describes the basin modeling in this case.  
Let us consider the following properties:  
$\phi_0$ = 0.36, 
$D$ = 0.03 cm, 
$a$ = 1.98 $\times$ 10$^{-22}$ mol/(cm$^2$ s), 
$b$ = 0.022 $^\circ$C$^{-1}$,
$\rho_1$ = 2.65 g/cm$^3$,  
$M_q$ = 60.09 g/mol,
$f$= 0.65, and $V$ = 1 cm$^3$, $\beta$ = 0.01 MPa$^{-1}$ and IGVi = 0.2.
The time step to solve equation (\ref{A1}) is $dt$ = 0.1 m.y.
Figure 13 shows the cement fraction and porosity, with and without compaction. In the latter case, the theory holds down to approximately 4.5 km depth, below which the porosity becomes negative and the whole pore space is filled with cement. 

Let us now consider two values of $f$, the amount of detrital quartz in the rock, i.e., $f$ = 0.4 (0.6 feldspar) and $f$ = 1.  Figure 14 indicates that nonquartz grains inhibit the cementation and 
the porosity loss. Using equation (\ref{A2}) indicates that the exponent in (\ref{A1}) is proportional to $f/D$ so 
that the effect of increasing the grain diameter is similar to that of decreasing the detrital quartz content, i.e., the amount of precipitated quartz
cement is less in coarse-grain sandstones, because of the reduced surface area, compared 
to fine-grain sandstones. 

The cementation only depends on the heating rate $H = G S$, so that different combinations of the geothermal gradient and sedimentation rate, but keeping the same heating rate, yield the same fraction of cement. However, it is not clear from equation (\ref{A1}) how $H$ affects the cementation, since it appears in several terms of the exponent. We consider three extreme values of $H$ = 0.4, 1.2 and 2 $^{\rm o}$C/m.y. Figure 15 indicates that $H$ greatly affects the process, where we assumed $f$ = 0.65 and $G$ = 30 $^{\rm o}$C/km. 
The plot shows that, for any given depth, the greatest
fraction of cement in the sandstone that had the most time to form is that corresponding to 
a heating rate of 0.4 $^{\rm o}$C/m.y., which at 3 km depth would be 225 m.y. old
and would have been above say, 80$^{\rm o}$C, for 25 m.y.. On the contrary, at 3 km depth the sandstone that
experienced a heating rate of 2 $^{\rm o}$C/m.y. would be only 45 m.y. old and would have been
above 80$^{\rm o}$C for only 5 m.y..

Next, we obtain the wave velocities. To compute the mineral HS averages in Models 1 and 3, the proportions of quartz (grain) and cement are $(1-\phi - \phi_p)/(1-\phi)$ and $\phi_p/(1-\phi)$, respectively. The three phases of Model 2 have proportions $1-\phi_p -\phi$ (grain), $\phi_p$ (cement) and $\phi$ (fluid or pore). 
The pore-aspect ratio in Model 2 is taken $\gamma$ = 0.108 and the bonding parameter in Model 4 is assumed to be $\delta$ = 100 MPa. 
Figure 16 shows the P- and S-wave velocities of the sandstones as a function of depth, corresponding to the four petro-elastical models. As can be observed, all the models are quite consistent with similar trends and 
values of the velocities, with Models 1 and 3 giving almost identical results. The research indicates that 
the use of simple models (e.g., Model 1) can be used in combination with the diagenesis-cementation process 
to make predictions. 

However, Models 2 and 4 have additional parameters, i.e., the pore aspect ratio ($\gamma$) (Model 2), and the maximum contact cement $\phi_{p0}$ and bonding parameters $\delta$ (Model 4). 
It can be shown that taking $\phi_{p0}$ = 0.1 (instead of 0.05) does not affect noticeably the results. 
On the other hand, the other two parameters ($\gamma$ and $\delta$) have an effect, as can be seen in Figure 17, where the velocities of Models 2 and 4 are also represented for $\gamma$ = 1 (spherical pores) and $\delta$ = 40 MPa (Gangi and Carlson, 1996). While the aspect ratio highly affects the velocity 
(e.g.,  Cheng et al., 2020), the initial bonding of the grains has a lesser influence. 

Next, we consider the more simple and complex models (1 and 4, respectively) and show the P-wave velocities for two values of the geothermal gradient $G$ = 20 and 30 $^{\rm o}$C/km ($H$ = 0.8 and 1.2 $^{\rm o}$C/m.y.) keeping the same sedimentation rate. The results are shown in Figure 18, where, as expected, a smaller temperature gradient (and heating rate) implies less diagenesis and cementation and lower velocities. Moreover, the two models predict the same trend and values of the velocities, indicating that the more simple Model 1 is suitable for predictions as Model 4, which is more complex from a mathematical point of view. 

Model 4 allows to change the location of contact cement from the grain contact to a uniform distribution on the grain surface. In this case, we use equation (\ref{dv21}) and assume 
$G$ = 30 $^{\rm o}$C/km, $\phi_{p0}$ = 0.05 and $\delta$ = 100 MPa. Figure 19 shows the velocities when the bonding cement is deposited at grain contacts and uniformly distributed. 
The differences are minimal in practice. 

The fact that the properties of the cement are close to those of quartz in Model 4 does not affect much the results. The CCT theory predicts that the type of cement has not a relevant influence on the 
the dry-rock moduli. Varying the cement stiffness one order of magnitude results
in only an about 15 \% increase in velocities, an
effect much smaller than that of increasing cement
content [Fig. 11 in Dvorkin et al. (1994)].

\section{Extensions of the models}

The smectite/illite petro-elastical model can be extended to the anisotropic case, as in Carcione and Avseth (2015), and to the presence of overpressure by using the pore-volume balance equations developed  by Carcione and Gangi (2000) for disequilibrium compaction, theories also implemented by Qin et al. (2019). 
Wangen (2000) developed an alternative theory of overpressure by cementation of sediments.
There is no overpressure if the released water caused by the smectite/illite reaction mostly escapes from the pore space (due in part to a relatively slow sedimentation rate), which is the case in the present work. 

Another type of diagenesis process that can easily be incorporated into the theory is the conversion of solid organic matter to hydrocarbons (Luo and Vasseur, 1996; Carcione et al., 2011), as mentioned in the introduction. Attenuation and velocity dispersion can be modeled with a theory based on physical principles (not developed yet, to our knowledge) or phenomenologically with the realistic assumption that cementation reduces seismic losses (Gei and Carcione, 2003), since attenuation decreases with depth and compaction. 
A completely different approach is the molecular dynamic modeling (Garcia and Medina, 2007), where 
at each step of the algorithm, cement is added at specific locations within the pores, in three different ways that model distinct origins and microgeometric features. 

\section{Conclusions}

We have developed a procedure to combine diagenesis and cementation processes for the formation of shales and sandstones from non-consolidated clay and sand, respectively, to different petro-elastical models to obtain the seismic velocities as a function of depth, pressure and temperature, on the basis of a linear basin modeling. 
The amount of illite in shales and quartz cementation in sandstones and resulting porosity loss and compaction affect the velocities, obtained with Gassmann-like equations, models based on oblate spheroid inclusions, and theories that assume a random packing of identical spherical grains (granular media). 

The proposed methods take into account in-situ properties such as the geothermal gradient, sedimentation and heating rates, elastic properties of the diagenetic material (cement), cement distribution in the pore space, pore aspect ratios, effective pressure and initial pre-stress conditions. All the models, based on different physical assumptions, predict the same increasing trend of velocity with burial history and similar values, indicating the robustness of the different theories, either based on the critical-porosity concept (e.g., Krief model) or an idealized shape of pores and grains to obtain the dry-rock stiffness moduli.  

\vspace{5mm}

%======================================================================

%{\bf Acknowledgments.}  aaaa

\newpage

\section*{REFERENCES} 

\vspace{-1cm}

\begin{verse}

\item
Allo, F., 2019,
Consolidating rock-physics classics:
A practical take on granular effective medium, 
The Leading Edge, 38(5), 334--340.

\item
Athy, L. F., 1930, Density, porosity, and compaction of
sedimentary rocks, AAPG Bulletin, 14, 1--24
 
\item
Avseth, P., and Lehocki, I., 2016,
Combining burial history and rock-physics modeling to
constrain AVO analysis during exploration, The Leading Edge, 
35(6), 528-534.
 
\item
Avseth, P., Mukerji, T., Mavko, G., and Dvorkin, J., 2010, Rock-physics diagnostics of depositional texture, diagenetic alterations, and reservoir heterogeneity in high-porosity siliciclastic sediments and rocks -- A review of selected models and suggested work flows, Geophysics, 75, 75A31--75A47. 

\item
Avseth, P., Skjei, N., and Mavko, G., 2016, 
Rock-physics modeling of stress sensitivity and 4D time
shifts in patchy cemented sandstones -- Application to
the Visund Field, North Sea,
The Leading Edge, 35(10), 868--878.

\item
Berryman , J. G., 1980, Long-wavelength propagation in composite elastic media II. Ellipsoidal
inclusions, J. Acoust. Soc. Am., 68, 1820--1831.

\item[]
Carcione, J. M., 2014, Wave Fields in Real Media.  Theory and 
numerical simulation of wave propagation  in anisotropic, 
anelastic, porous and electromagnetic media, 3rd edition, 
revised and extended, Elsevier Science.

\item[]
Carcione, J. M., and Avseth, P., 2015, Rock-physics templates for clay-rich source rocks, Geophysics, 80, D481--D500.

\item
Carcione, J. M., and Gangi, A., 2000, Non-equilibrium compaction and abnormal pore-fluid pressures: effects on seismic attributes, Geophys. Prosp., 48, 521--537.

\item
Carcione, J. M., Farina, B., Poletto, F., Qadrouh, A. N., and Cheng, W., 2020, Seismic
attenuation in partially molten rocks, Physics of the Earth and Planetary Interiors,
https://doi.org/10.1016/j.pepi.2020.106568

\item[]
Carcione, J. M., Helle, H. B., and Avseth, P., 2011, Source-rock seismic-velocity models,
Gassmann versus Backus: Geophysics, 76, N37-N45.

\item[]
Carcione, J. M., Helle, H. B., Santos, J. E., and Ravazzoli, C. L., 2005, A constitutive
equations and generalized Gassmann modulus for multi-mineral porous media,
Geophysics, 70, N17-N26.

\item
Carcione, J. M., Landr\o, M.,  Gangi, A. F., and Cavallini, F., 2007,
Determining the dilation factor in 4D monitoring of compacting reservoirs
by rock-physics models, Geophys. Prosp., 55, 793--804. 

\item
Carcione, J. M., Poletto, F., and Farina, B., 2018, 
The Burgers/squirt-flow seismic model of the crust and mantle, 
Physics of the Earth and Planetary Interiors, 274, 14--22. 

\item
Cheng, W., Carcione, J. M., Qadrouh, A., Alajmi, M., and Ba, J., 2020, Rock anelasticity, pore geometry and the Biot-Gardner effect, Rock Mechanics and Rock Engineering, \url{https://doi.org/10.1007/s00603-020-02155-7}.

\item
Draege, A., Jakobsen, M., and Johansen, T. A., 2006, 
Rock physics modelling of shale diagenesis, 
Petroleum Geoscience, 12, 49--57.

\item
Dvorkin, J., Berryman, J., and Nur, A, 1999, Elastic moduli of cemented sphere
packs, Mechanics of Materials, 31, 461--469.

\item
Dvorkin, J., Nur, A., and Yin, H., 1994, Effective properties of
cemented granular material, Mechanics of Materials 18, 351--366.

\item
Dvorkin, J., and Nur, A., 1996, 
Elasticity of high-porosity sandstones: Theory for two North Sea data sets, 
Geophysics, 61, 1363--1370. 

\item
Gangi, A. F., and Carlson, R. L., 1996, An asperity-deformation model for effective pressure,Tectonophysics, 256, 241--251. 

\item
Garcia, X., and Medina, E., 2007, 
Acoustic response of cemented granular sedimentary rocks: Molecular dynamics modeling, 
Physical Review E, 75, 061308 2007.

\item[]
Gei, D., and J. M. Carcione, 2003, Acoustic properties of sediments saturated with gas
hydrate, free gas and water: Geophys. Prosp., 51, 141-157.

\item
Goffe, W. L., Ferrier, G. D., and Rogers, J., 1994, Global optimization of statistical functions with simulated annealing, Journal of Econometrics, 60(1-2), 65--99. 
      
\item
Gurevich, B. and Carcione, J. M., 2000, Gassmann modeling of acoustic properties of
sand/clay mixtures, Pure and Applied Geophysics, 157, 811--827.

\item
Hashin, Z., and Shtrikman, S., 1963, A variational approach to the theory of the elastic
behaviour of multiphase materials, Journal of the Mechanics and Physics of Solids,
11, 127--140.

\item
Hertz, H., 1895, 
Gesammelte Werke, Band I, Schriften Vermischten Inhalts, Leipzig. 

\item
Keller, W. D., Reynolds, R. C., and Inoue, A., 1986, Morphology of clay minerals in the smectite to illite conversion series by scanning electron microscopy, Clays \& Clay Minerals, 34, 187--197. 

\item[]
Krief, M., J. Garat, J. Stellingwerff, and J. Ventre, 1990, 
A petrophysical interpretation using the velocities of P and S waves 
(full waveform sonic):
The log Analyst, 31, 355-369. 

\item
Lander, R. H., and O. Walderhaug, 1999, Predicting porosity through simulating sandstone compaction and quartz cementation, AAPG Bulletin, 83, 433--449.

\item
Luo, X., Vasseur, G., 1996, Geopressuring mechanism of organic matter cracking: Numerical modeling, AAPG Bulletin, 80, 856--874

\item
Man, C.-S., and M. Huang, 2011, A simple explicit formula for the Voigt-Reuss-Hill average of elastic polycrystals with arbitrary crystal and texture symmetries, Journal of Elasticity, 105, 29--48. 

\item[]
Mavko, G., T. Mukerji, and J. Dvorkin, 2009, The rock physics handbook: tools for
seismic analysis in porous media, Cambridge Univ. Press.

\item
Mindlin, R. D., 1949,
Compliance of elastic bodies in contact,
J. Appl. Mech., 16, 259-268. 

\item[]
Mondol, N. H., J. Jahren, K. Bj\o rlykke, and I. Brevik, 2008, 
Elastic properties of clay minerals, 
The Leading Edge,  27(6), 758-770.

\item
Paxton, S. T.,  Szabo, J. O., Ajdukiewicz, J. M., 
and Klimentidis, R. E., 2002, Construction of an intergranular volume compaction curve for evaluating and predicting compaction and porosity loss in rigid-grain sandstone reservoirs, 
AAPG Bulletin, 86, 2047--2067. 

\item
Perry, E., and Hower, J., 1970, Burial diagenesis in Gulf Coast pelitic sediments, Clays \& Clay Minerals 18, 165--177

\item
Picotti, S., Carcione, J. M., and Ba, J., 2018, 
Rock-physics templates based on seismic $Q$, Geophysics, 84, MR13--MR23. 

\item[]
Pinna, G., J. M. Carcione, and F. Poletto, 2011, Kerogen to oil conversion in source
rocks. Pore-pressure build-up and effects on seismic velocities: J. Appl. Geophy., 74, 229--235.

\item[]
Pytte, A. M., and R. C. Reynolds, 1989, The thermal transformation of smectite
to illite. In N. D. Naeser and T. H. McCulloh, editors, Thermal History of
Sedimentary Basins: Methods and Case Histories, pages 133--140. Springer--Verlag.

\item
Qadrouh, A. N., Carcione, J. M., Alajmi, M., and Ba, J., 2020, 
Bounds and averages of seismic $Q$, Stud. Geophys. Geod., 
https://doi.org/10.1007/s11200-019-1247-y. 

\item
Qin, X., Han, D.-H., and Zhao, L., 2019, 
Elastic characteristics of overpressure due to smectite-to-illite
transition based on micro-mechanism analysis, 
Geophysics, 84, 1JA-Z21. 

\item
Rieke III, H. H., and Chilingarian, G. V., 1974, Compaction of
argillaceous sediments. Elsevier, New York.

\item 
Sadikh-Zadeh, L. A., 2006, 
Prediction of sandstone porosity through quantitative
estimation of its mechanical compaction during
lithogenesis (Bibieybat Field, South Caspian Basin),
Natural Resources Research, 15, DOI: 10.1007/s11053-006-9008-3

\item[]
Scotchman, I. C., 1987, Clay diagenesis in the Kimmeridge Clay Formation, onshore UK, and its relation to
organic maturation: Mineral. Mag., 51, 535-551.

\item
Sippel, R. F., 1968, Sandstone petrology, evidence from luminescence
petrography, Journal of Sedimentary Petrology, 38, 530--554.

\item
Stephenson, L. P., Plumley, W. J., and Palciauskas, V. V., 1992, A
model for sandstone compaction by grain interpenetration:
Journal of Sedimentary Petrology, 62, 11--22.

\item[]
Thyberg, B. I., J. Jahren, T. Winje, K. Bj\o rlykke, and J. I. Faleide, 2009. From mud to shale:
rock stiffening by micro-quartz cementation: First Break, 27, 27-33.

\item
Walderhaug, O.,1996, Kinetic modelling of quartz cementation and porosity
loss in deeply buried sandstone reservoirs. AAPG Bulletin, 80, 731--745.

\item
Walderhaug, O., 2000, Modeling quartz cementation and porosity in Middle Jurassic Brent Group sandstones of the Kvitebj\o rn field, northern North Sea, AAPG Bulletin, 84, 1325--1339. 

\item
Walton, K., 1987, 
The effective elastic moduli of a random packing of spheres, 
J. Mech. Phys. Solids, 35, 213--226. 

\item
Wangen, M., 2000, 
Generation of overpressure by cementation of pore space in sedimentary rock, 
Geophys. J. Int., 143,  608--620. 

\item
Weyl, P. K., 1959, Pressure solution and the force of crystallization -- a phenomenological theory, J. Geophys. 
Res., 64, 2001--2025.

\end{verse}

\newpage

\section*{Statements and Declarations}

The authors declare that no funds, grants or other support were received during the preparation of this manuscript, and have no relevant financial or non-financial interests to disclose.

All authors contributed to the study conception and design. The manuscript was written by J. M. Carcione, and 
D. Gei and S. Picotti checked the shale and sandstone sections, respectively. A. N. Qadrouh, M. Alajmi and J. Ba contributed to the review of the literature,  data collection and analysis.  
All authors commented on previous versions of the manuscript and approved the final manuscript.

\newpage

\numberwithin{equation}{section}

\begin{appendices}

\section{Smectite/illite conversion kinetic model}

The transformation of smectite to illite is probably the most important clay mineral reaction in 
sedimentary rocks (Perry and Hower, 1970). 
Pytte and Reynolds (1989) propose a model for the smectite/illite ratio $r$ based on the $n$th-order Arrhenius-type reaction
\begin{equation} \label{si1}
\frac{\partial r}{\partial t} = - r^n c \exp [-E / R T(t)] , 
\end{equation}
where $E$ is the activation energy, 
$R$ = 1.986 cal/( mol $^{\rm o}$K ) is the gas constant, $c$ is a constant and $T(t)$ is the absolute temperature.   

 The illite/smectite ratio in percent is 100 ($1 - r$). The solution of equation \ref{si1} is 
\begin{equation} \label{si2}
r (t) = m^{-1/m} \left\{ \frac{r_0^{-m}}{m} + \frac{c}{H} \left[ \frac{E}{R} \left[ {\rm Ei}(x) - {\rm Ei}(x_0) \right] + T \exp (x) - T_0 \exp (x_0)  \right] \right\}^{- 1/m}  , 
\end{equation}
where $r_0$ is the initial ratio, $m = n-1$, Ei ($x$) is the exponential integral,
\begin{equation} \label{si3}
x= - \frac{E}{R T} , \ \ \ x_0 = - \frac{E}{R T_0} , 
\end{equation}
and the dependence on the deposition time is given in the absolute temperature [see equation (\ref{tem})]. To compute (\ref{si2}), we use the property Ei($x$) = $-$ ${\rm E}_1(-x)$. 

\section{Quartz cement precipitated in sandstones and compaction}

It is assumed that dissolved silica is the source of cement (quartz dissolution) at stylolites or at grain contacts containing clay or
mica, and diffuses short distances to the sandstone volume, so that no
quartz dissolution or grain interpenetration takes
place within this volume. Following Walderhaug (1996), the amount of quartz cement (cm$^3$)
precipitated in a 1 cm$^3$ of sandstone at time $t_{n+1}$ is 
\begin{equation} \label{A1}
\phi_p^{n+1} = \phi_0 - (\phi_0 - \phi_p^n) 
\exp \left[ -
\frac{M_q a A_0}{\rho_q \phi_0 b H \ln 10} \left( 10^{b T_{n+1}}-10^{b T_n} \right) 
\right], \ \ \ T_n = H t_n + T_0, 
\end{equation}
where $\phi_p^n$ is the amount of quartz cement present at time $t_n$, 
$\phi_0$ is the initial porosity,
$a$ and $b$ are constants, which have units of mol/(cm$^2$ s) and $^\circ$C$^{-1}$,
respectively, 
$H$ is the heating rate, 
$\rho_q$ is the quartz density, 
$M_q$ is the molar mass of quartz, and
\begin{equation} \label{A2}
A_0 = \frac{6 f V}{D}
\end{equation}
is the initial quartz surface area, where $D$ is the grain diameter, $f$ is the fraction of detrital 
quartz in the rock and $V$  is a unit volume (1 m$^3$ if $D$ is given in meters). 
Walderhaug (1996) shows examples with 65 and 50 \% detrital quartzite and the rest is 
feldspar.

In the first time step, $t_0$ = 0, $\phi_p^0$ = 0, $T_1 = H dt + T_0$ and 
\begin{equation} \label{A3}
\frac{\phi_p^{1}}{\phi_0} = 1 - \exp \left[ -
\frac{M_q a A_0}{\rho \phi_0 b H \ln 10} ( 10^{b T_{1}}-10^{b T_0} ) 
\right],
\end{equation}
where $dt$ is the time step. 

The porosity varies as 
\begin{equation} \label{A4}
\phi = \phi_0 - \phi_p ,
\end{equation}
and the surface area as
\begin{equation} \label{A5}
A = A_0 \frac{\phi}{\phi_0}. 
\end{equation}
The effect of clay coating is to reduce the precipitation and is modeled as 
\begin{equation} \label{A6}
A_0 \rightarrow (1 - {\cal C} ) A_0, 
\end{equation}
where ${\cal C}$ is a clay factor. ${\cal C}$ = 1 implies zero surface area and no precipitation. The clay factor affects the surface area in the same manner as the fraction of 
detrital quartz $f$ and reciprocally with respect to the grain diameter $D$.  

Equation (\ref{A1}) assumes a constant heating rate. An algorithm for a variable heating rate is gieven by Eq. (5) in Walderhaug (1996). 

Porosity reduction by compaction is based on the intergranular volume loss index (IGV)
\begin{equation} \label{A7}
\mbox{IGV} = \mbox{IGVi} + (\phi_0 + m_0 - {\rm IGVi} ) \exp ( - \beta p_e )
\end{equation}
(Lander and Walderhaug, 1999), where IGVi is the value at infinite effective pressure, $m_0$ is the initial matrix volume fraction in the pore space, 
and $\beta$ is a constant. If $p_e$ = 0, IGV = $\phi_0 + m_0$, the maximum value. For 
IGVi = $m_0$ = 0, we obtain Athy equation (\ref{Athy}), while for 
IGVi = $\phi_0$ and $m_0$ = 0, we obtain IGV = $\phi_0$. 
A detailed pictorial explanation of the IGV index can be found in 
Paxton et al. (2002). IGVi and $\beta$ are obtained by fitting experimental data. 

In the absence of matrix material, the porosity is given by 
\begin{equation} \label{A8}
\phi = \mbox{IGV} - \phi_p 
\end{equation}
(e.g., Sadikh-Zadeh, 2006). 
Equation (\ref{A4}) is obtained for $p_e$ = 0, neglecting the pressure effects that lead to 
compaction. Actually, there is no particular theory behind equation (\ref{A7}), which is an empirical equation of the type used by Athy (1930) ninety years ago, of the form $\phi = \phi_0 \exp (- \beta^\prime z)$, where $\beta^\prime = g (\bar \rho - \bar \rho_w ) \beta$. 

\section{Hashin and Shtrikman bounds and averages}

Let us denote the solid bulk and shear moduli by $K_i$ and $\mu_i$, respectively. A two-solid composite, with 
no restriction on the shape of the two phases, has stiffness bounds given by the 
Hashin and Shtrikman (1963) equations, 
\begin{equation} \label{HSK-}
K_{\rm HS}^\pm = K_1 +  
\frac{\beta_2}{(K_2 - K_1)^{-1} + \beta_1 \left( K_1 + \myfrac{4}{3} \mu_\beta \right)^{-1}}   
\end{equation}
and 
\begin{equation} \label{HSK+}
\mu_{\rm HS}^\pm = \mu_1 +  
\frac{\beta_2}{(\mu_2 - \mu_1)^{-1} + \beta_1 \left[ \mu_1 +  \myfrac{\mu_\beta}{6}
\left(  \myfrac{9 K_\beta + 8 \mu_\beta}{K_\beta + 2 \mu_\beta} \right) 
\right]^{-1}}  ,
\end{equation}
where $\beta_1$ and $\beta_2$ are the fractions of solid 1 and 2 ($\beta_1+\beta_2$ = 1), and
$\beta_1 = \phi_1 /(1-\phi)$.  
We obtain the upper bounds when $K_\beta$ and $\mu_\beta$ are the maximum bulk and shear moduli of the single components, and the lower bounds 
when these quantities are the corresponding minimum moduli, i.e., we have the upper bound if 1 is the stiffer medium and the lower bound is obtained if 1 is the softer medium (Mavko et al., 2009). 

The arithmetic averages of the bounds are frequently used to obtain the bulk and shear moduli of a mineral mixture, i.e., 
\begin{equation} \label{HSav}
K_s = \frac{1}{2} (K_{\rm HS}^+ + K_{\rm HS}^- ) , \ \ \ 
\mu_s = \frac{1}{2} (\mu_{\rm HS}^+ + \mu_{\rm HS}^- ) . \ \ \ 
\end{equation}

\section{The Hertz-Mindlin model}

We model pressure dependence in Model 4 by using the Hertz-Mindlin (HM) contact theory, which considers spherical grains. 
The stresses are calculated in terms of the
strains by considering the random packing 
of spheres as an effective medium that exerts a mean-field force 
(as given by the contact Hertzian theory) on a single representative grain (Hertz, 1895; Mindlin, 1949; Walton, 1987; Mavko et al., 2009). 

We modify the Hertz-Mindlin model by replacing the effective pressure $p_e$, by the augmented value, $p_e + \delta$, following Gangi and Carlson (1996), assuming that at the beginning of the burial process the grains are not ``floating" in the fluid, but there is an initial level of bonding between grains determined by $\delta$ (a pre-stress condition). A HM model with augmented effective pressure is given in Carcione et al. (2007) to model bonded grains at $p_e$ = 0. 

Then, the bulk and shear (uncemented) moduli at the critical porosity $\phi_0$ are given by 
\begin{equation}\label{KHM}
K_{u} = \left[ \frac{{C}^{2} (1 - \phi_0)^{2} \mu_{1}^{2} (p_{e} + \delta) } {18 \pi^{2} (1 - \nu_{1})^{2}} \right]^{1/3}
\end{equation}
and 
\begin{equation}\label{muHM}
\mu_{u} = \frac{3 (5 - 4 \nu_1)}{5 (2 - \nu_1)} K_{u} ,
\end{equation}
where $\mu_1$ is the shear modulus of the grains, $\nu_1$ is the Poisson ratio of the grains and ${C}$ is the average number of contacts per spherical grain. 

\end{appendices}

\newpage

\section{Tables} \label{J}

%======================================================================
\begin{center}
{\bf Table 1. Material Properties.} \\                  
\end{center}
%====================================================================                                                                                
%---------                                                                    
\[
\begin{tabular}{|c|c|c|c|c|c|c|} 
\hline
{\rm property} & smectite & illite & quartz & cement & water \\  
\hline
$K$ (GPa) & 9 & 33 & 40 & 35 & 2.2 \\
$\mu$ (GPa) & 6 & 28 & 39 & 30  & 0 \\ 
$\rho$ (g/cm$^3$) & 2.25 & 2.8 & 2.65 & 2.60  & 1.04 \\
\hline
\end{tabular}                   
\]
\normalsize

\newpage

\begin{figure}
\hspace{-2cm}\includegraphics[width=15cm]{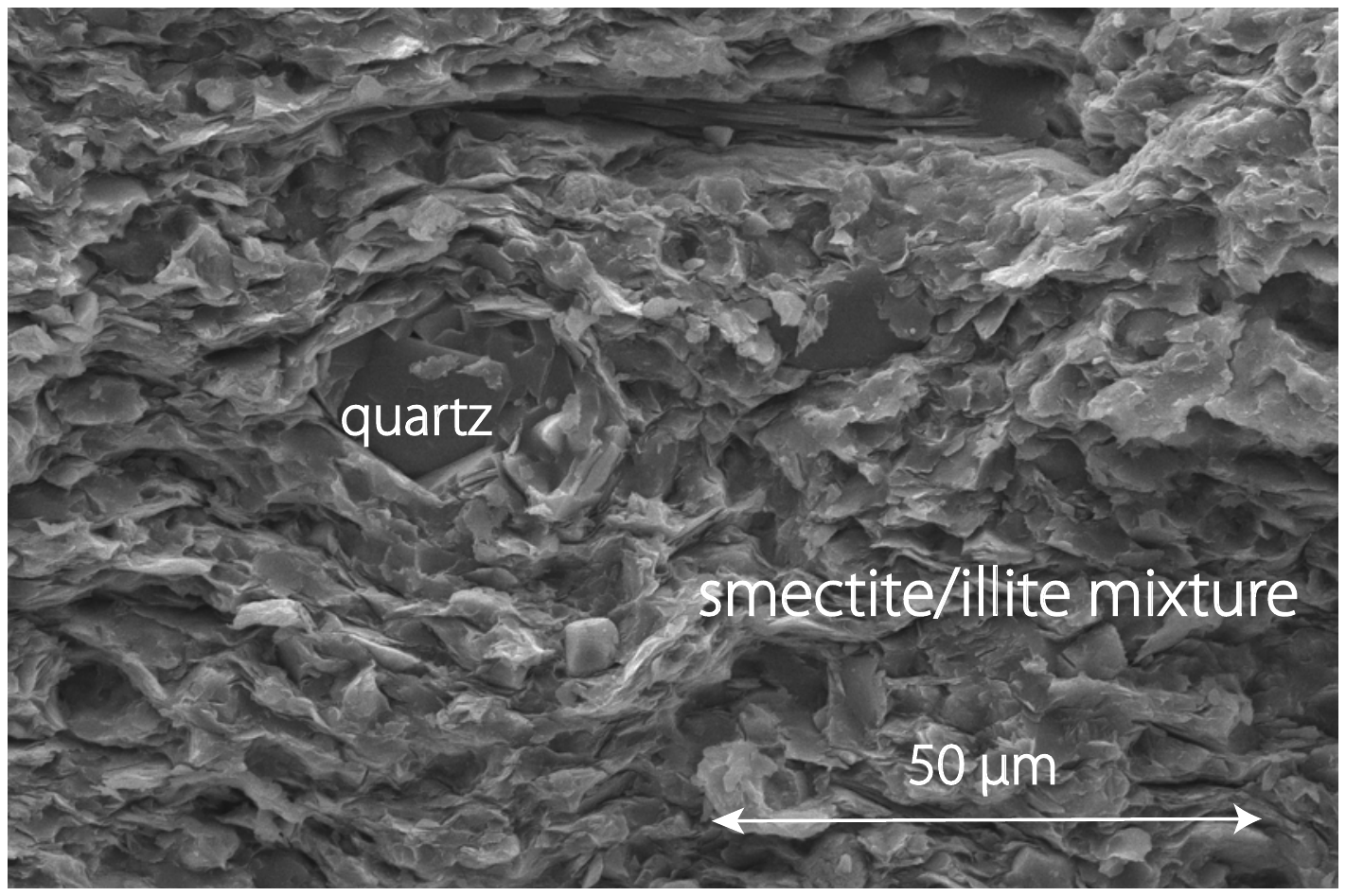}
\vspace{-7cm}
\caption{
Scanning electron micrographs (SEM) of a shale showing a texture corresponding to a   
smectite/illite mixture (Sichuan Basin, China; depth: 3.3 km).
}
\end{figure}

\begin{figure}
\hspace{-3cm}\includegraphics[width=16cm]{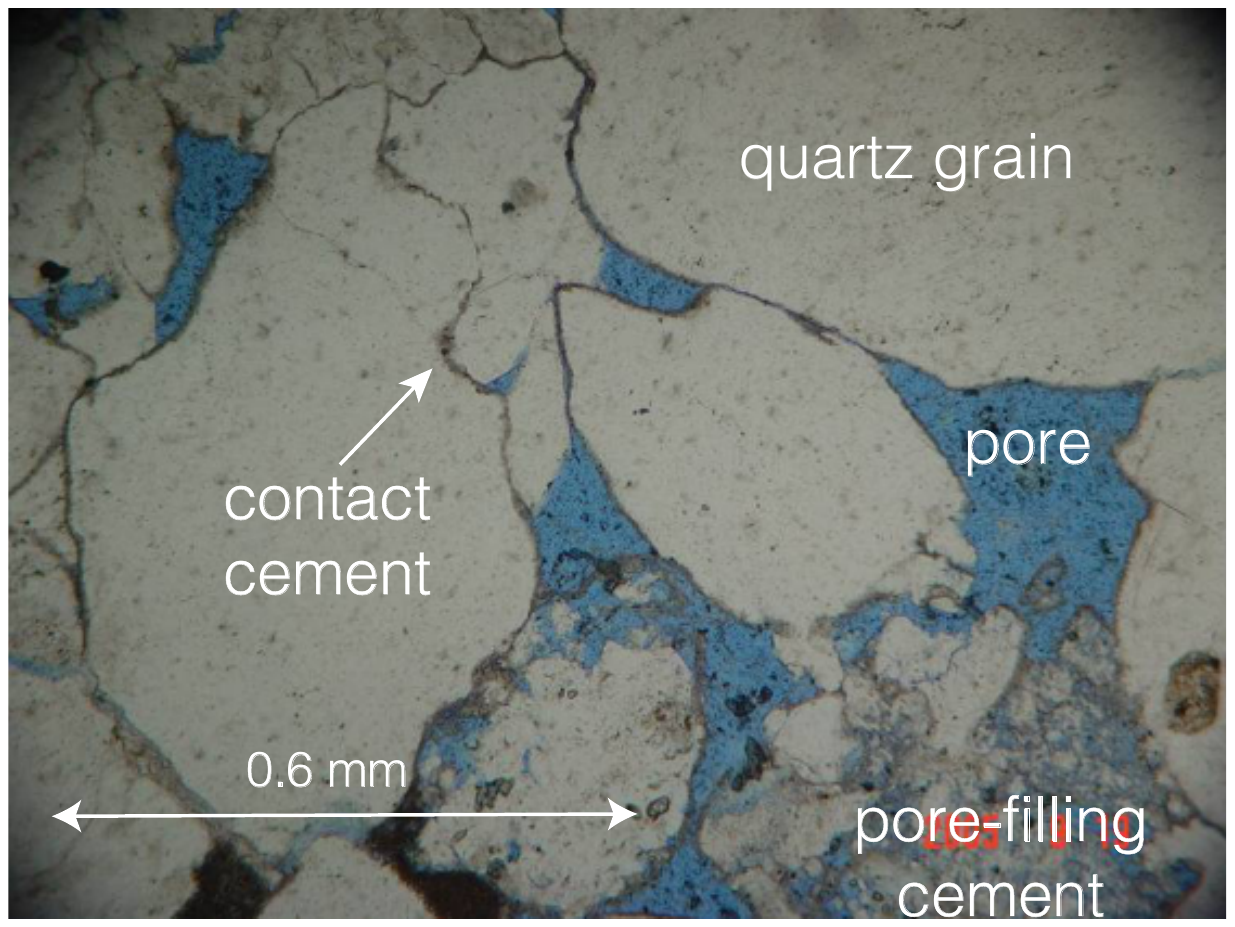}
\vspace{-7cm}
\caption{
Petrographic thin section of a sandstone showing the mineral composition 
(Sichuan Basin, China; depth: 2 km).
}
\end{figure}

\begin{figure}
\hspace{-3cm}\includegraphics[width=18cm]{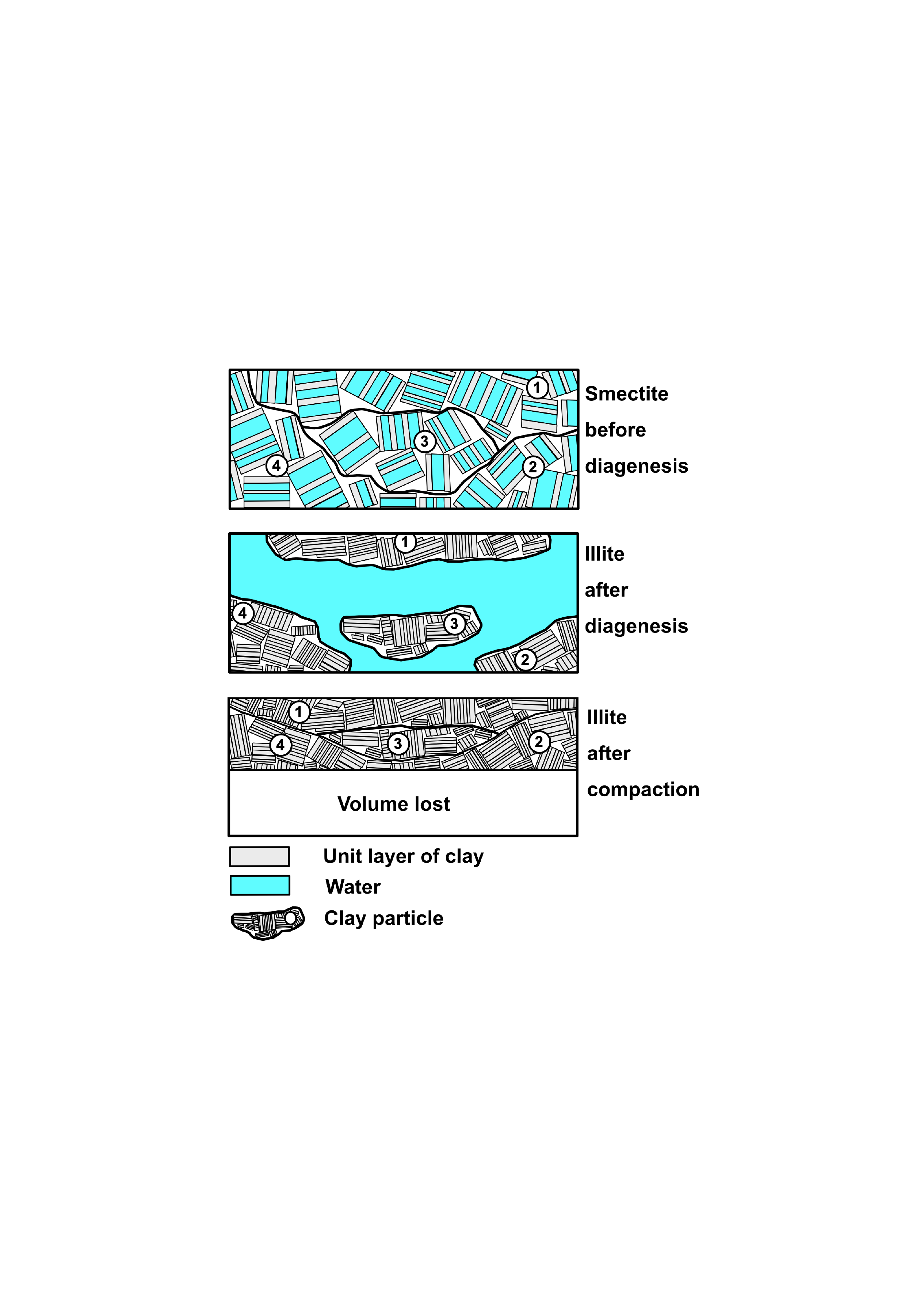}
\vspace{-7cm}
\caption{
Clay diagenesis. Smectite/illite conversion with release of 
bound water. Clays consist of bound water at the time of
deposition. Free water increases with burial with the
consequent release of bound water. Then, free water is
squeezed out of the original volume (modified from Rieke III and Chilingarian, 1974, Fig. 57). 
}
\end{figure}

\begin{figure}
\includegraphics[width=14cm]{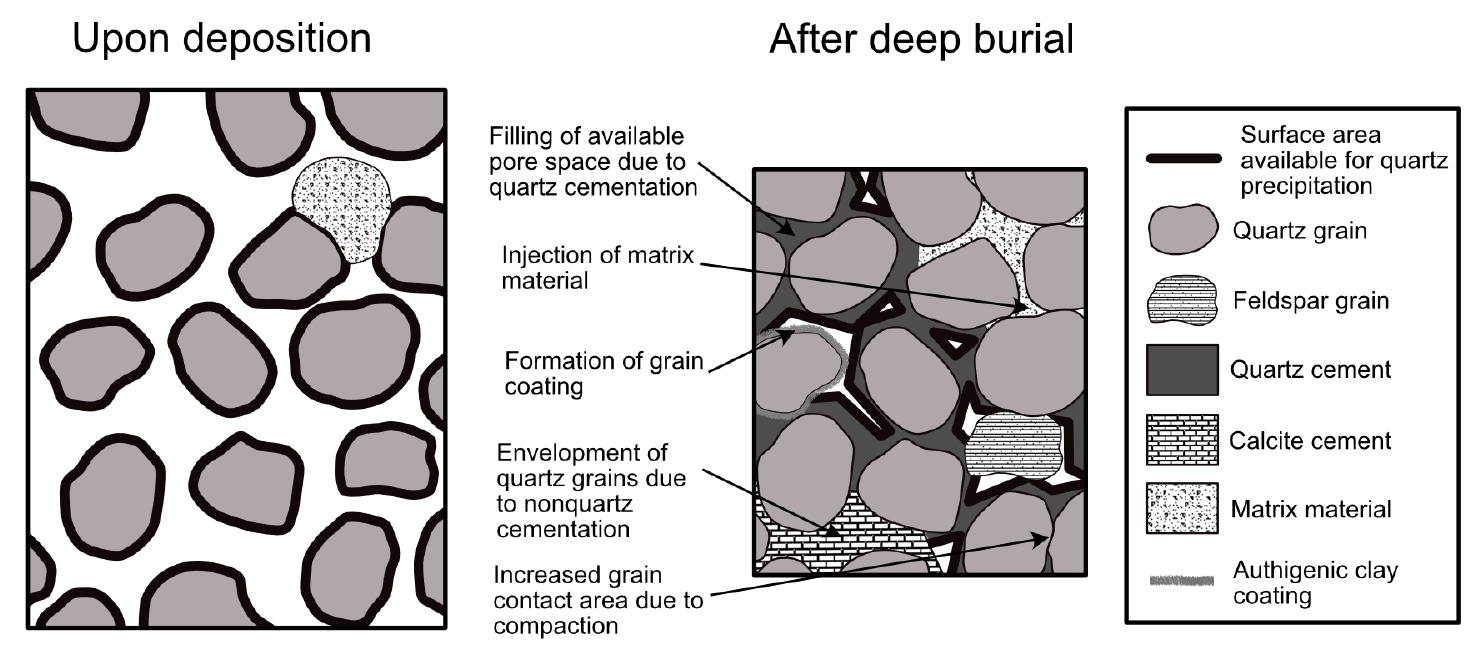}
\caption{
Sandstone diagenesis and cementation.
Compaction reduces quartz surface area by increasing grain contact
area, as well as by the injection of matrix material
into the pore space. Cementation causes surface
area reduction when quartz grains are encased by
pore-filling cements (modified from Lander and Walderhaug, 1999).
}
\end{figure}

\begin{figure}
\includegraphics[width=8cm]{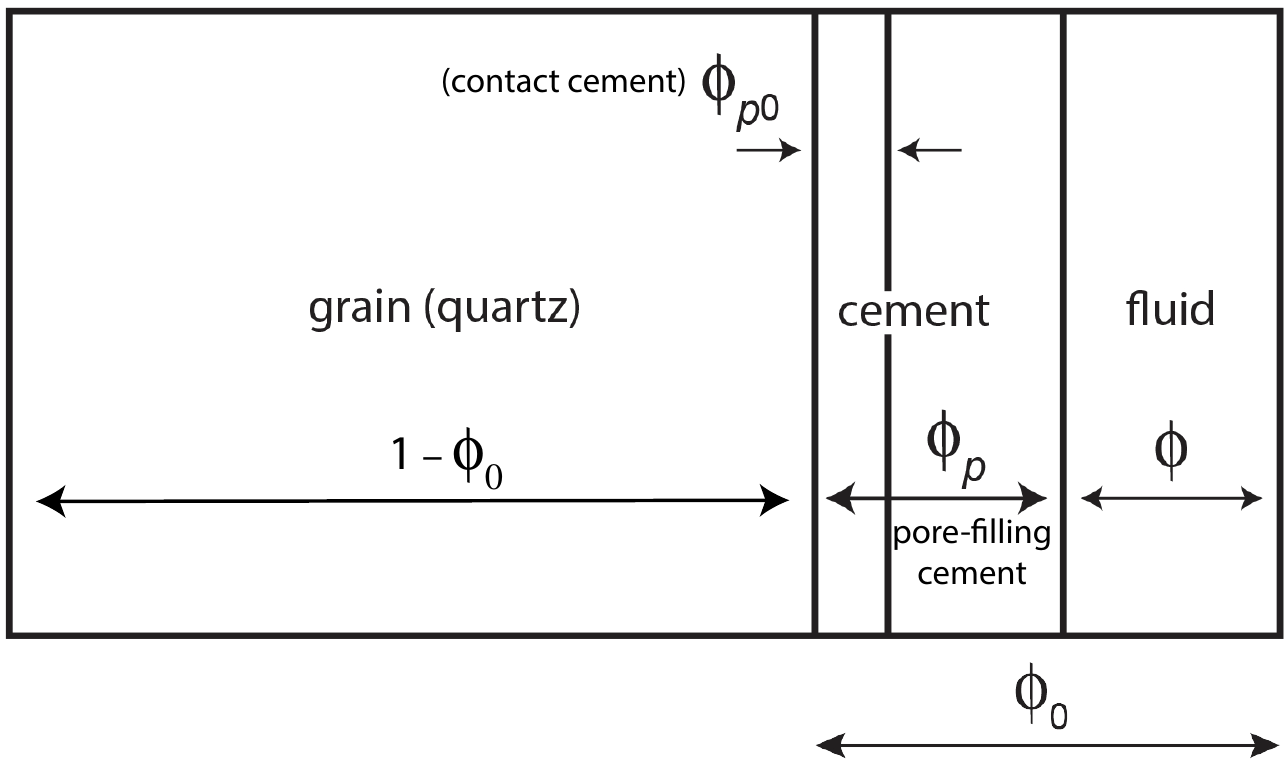}
\caption{
Model 4. Sandstone after cementation: $\phi_{p0}$ is the maximum cement fraction bonded to the grains, $\phi_p$ is the total volume fraction of cement, and $\phi_p - \phi_{p0}$ is the pore-filling cement fraction.}
\end{figure}

\begin{figure}
\includegraphics[width=6cm]{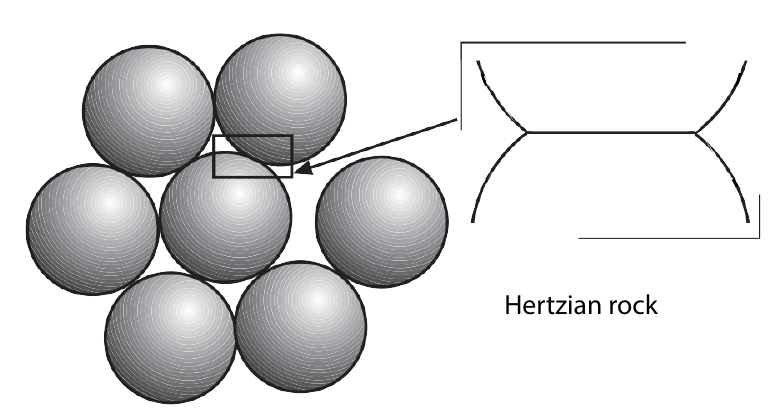}
\caption{
A random packing of spheres corresponding to the Hertz-Mindlin model, where the contact area between the grains depends on the effective pressure.}
\end{figure}

\begin{figure}
\includegraphics[width=11cm]{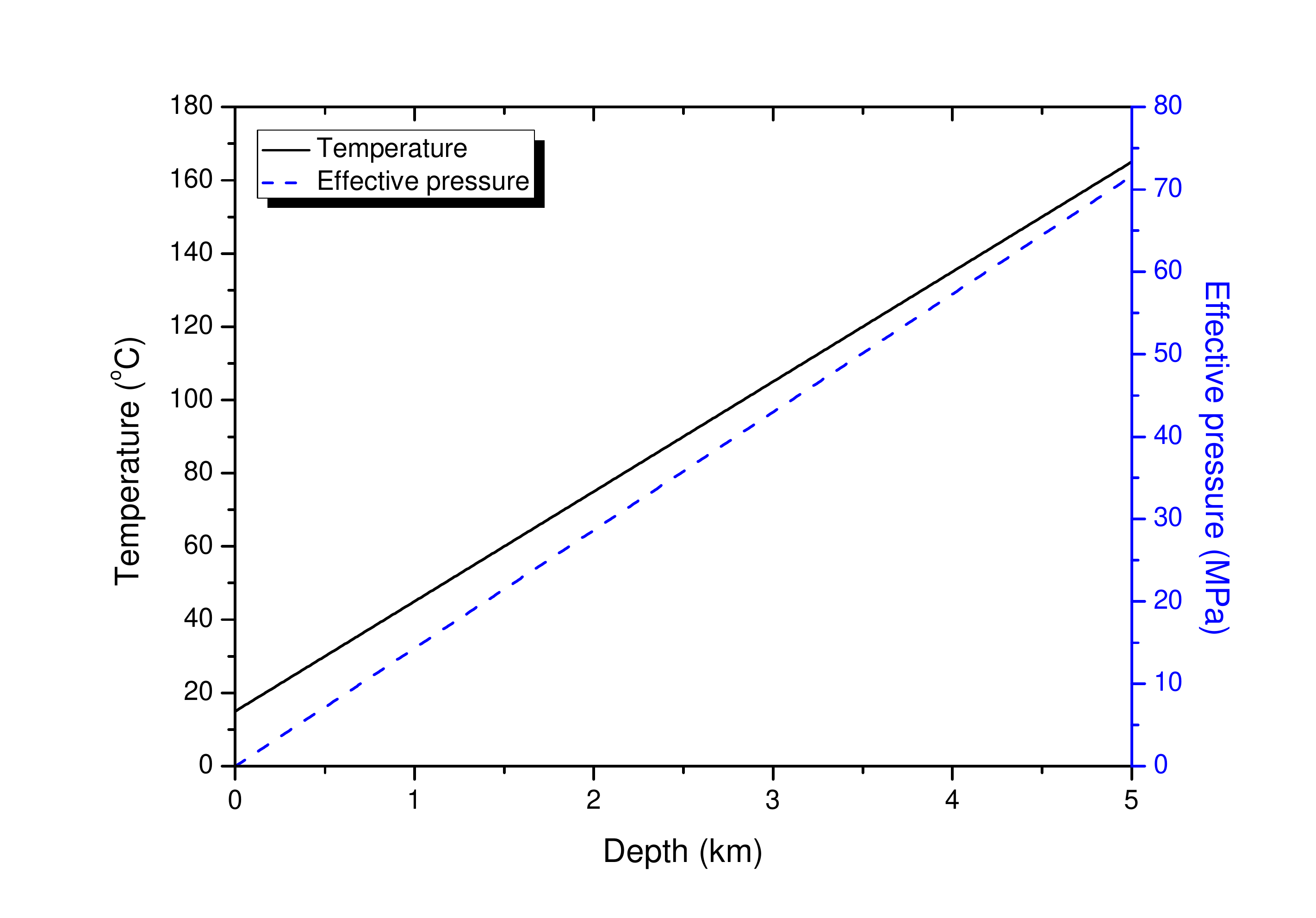}
\caption{
Temperature-pressure-depth relation of the linear basin modeling.}
\end{figure}

\begin{figure}
\includegraphics[width=10cm]{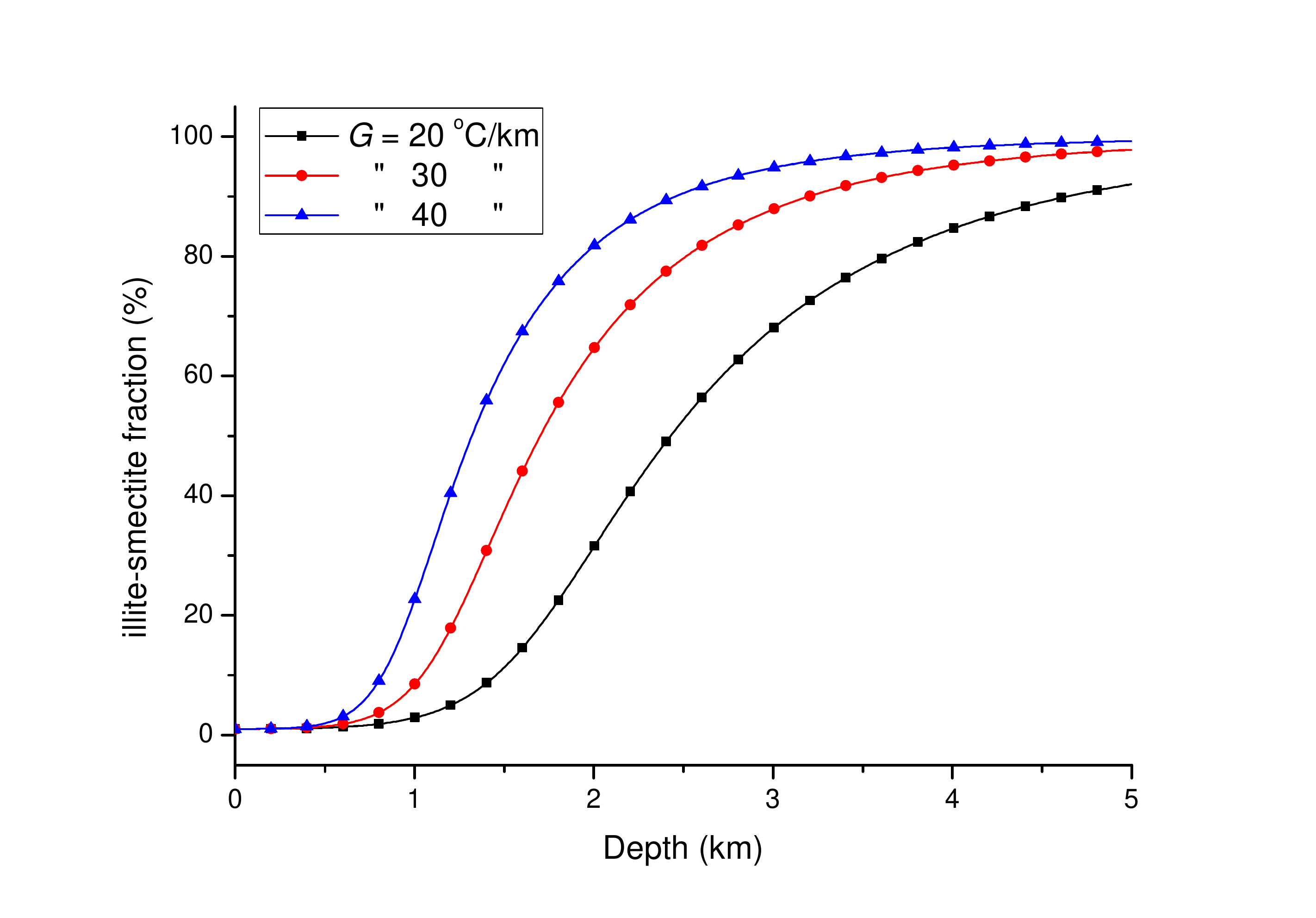}
\caption{
Illite-smectite fraction as a function of depth for three values of the geothermal gradient.}
\end{figure}

\begin{figure}
\includegraphics[width=10cm]{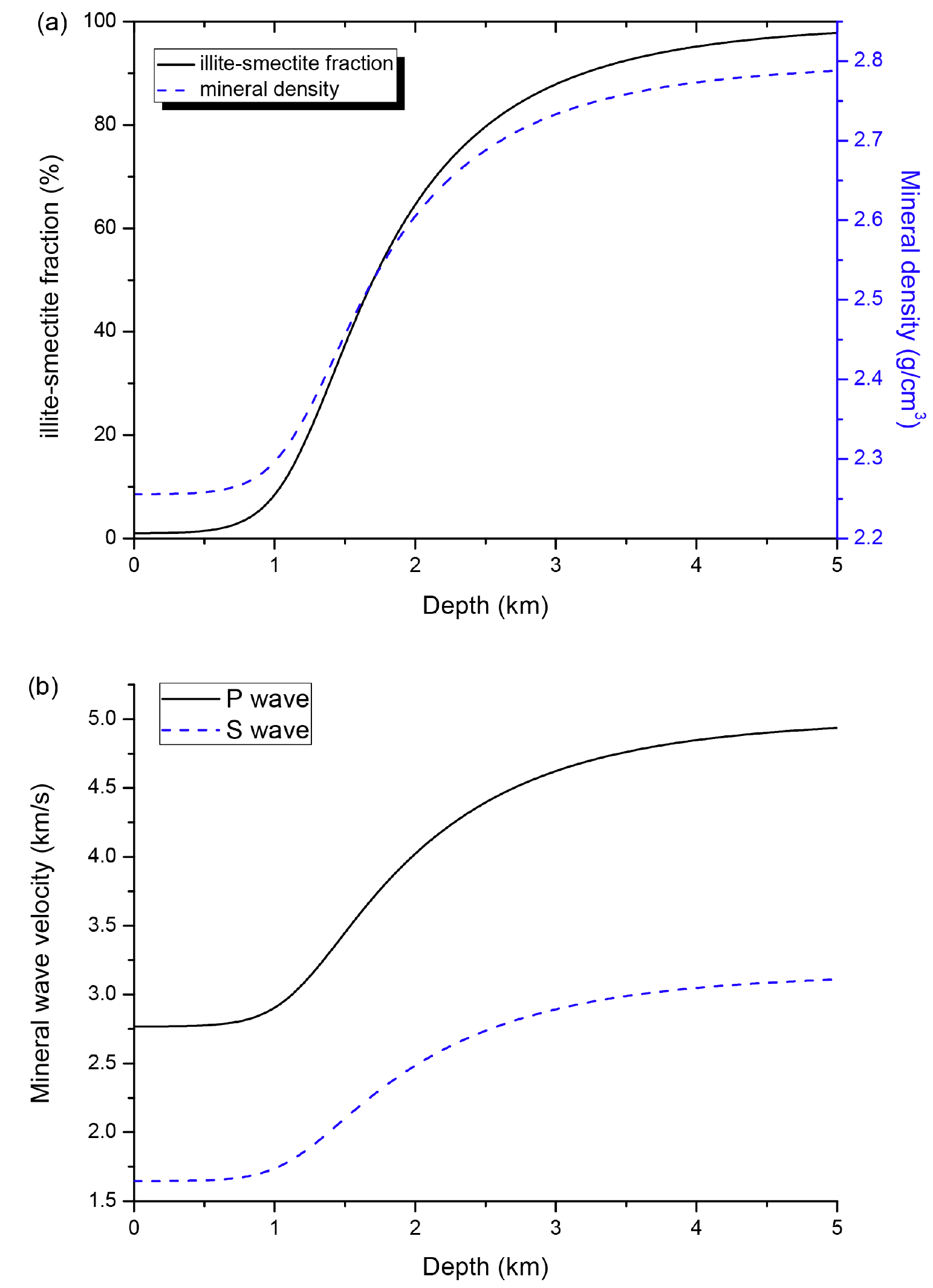}
\caption{
Illite-smectite fraction and density (a), and wave velocities of the mineral mixture as a function of depth.}
\end{figure}

\begin{figure}
\includegraphics[width=10cm]{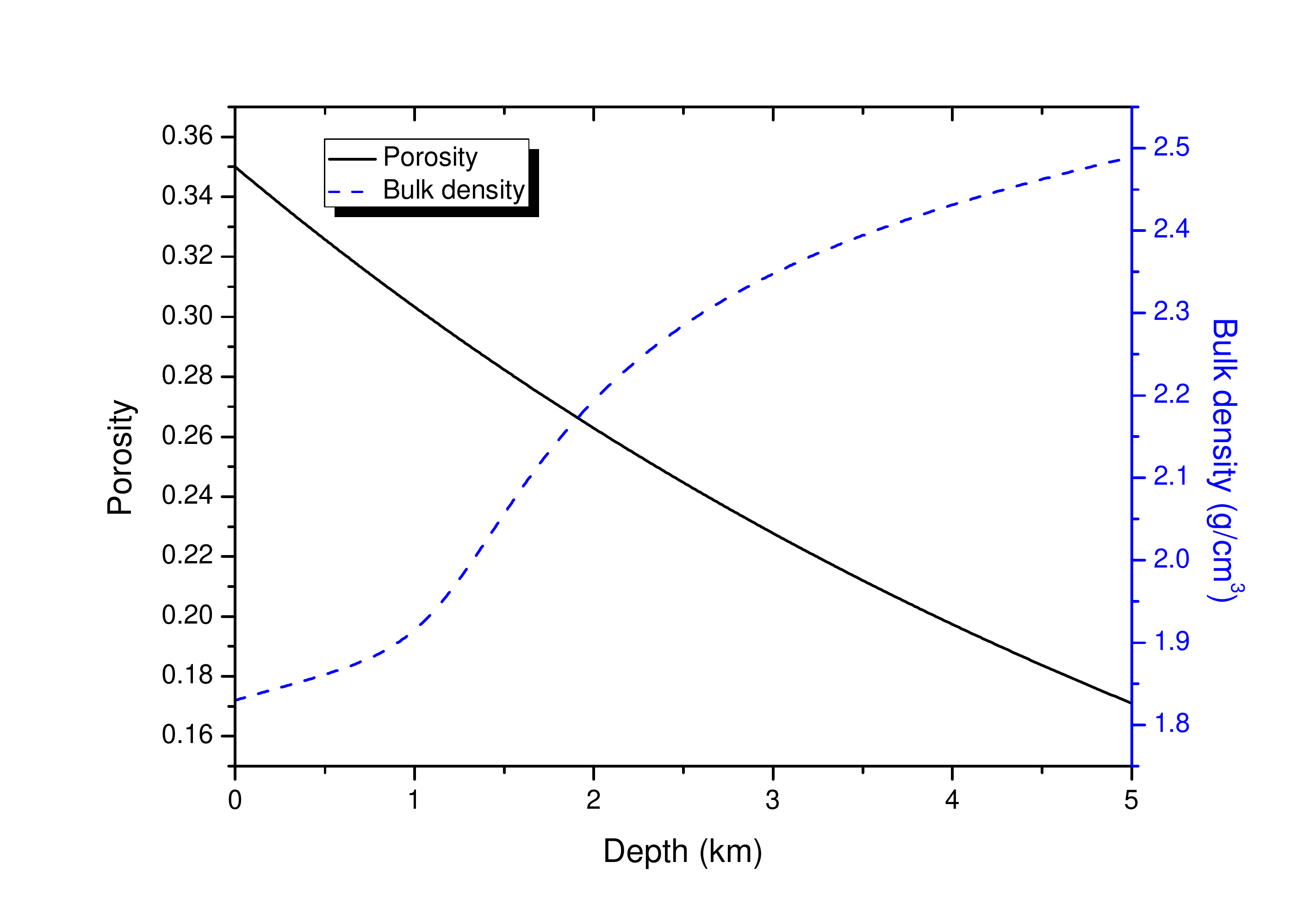}
\caption{
Shale porosity and bulk density as a function of depth.}
\end{figure}

\begin{figure}
\includegraphics[width=10cm]{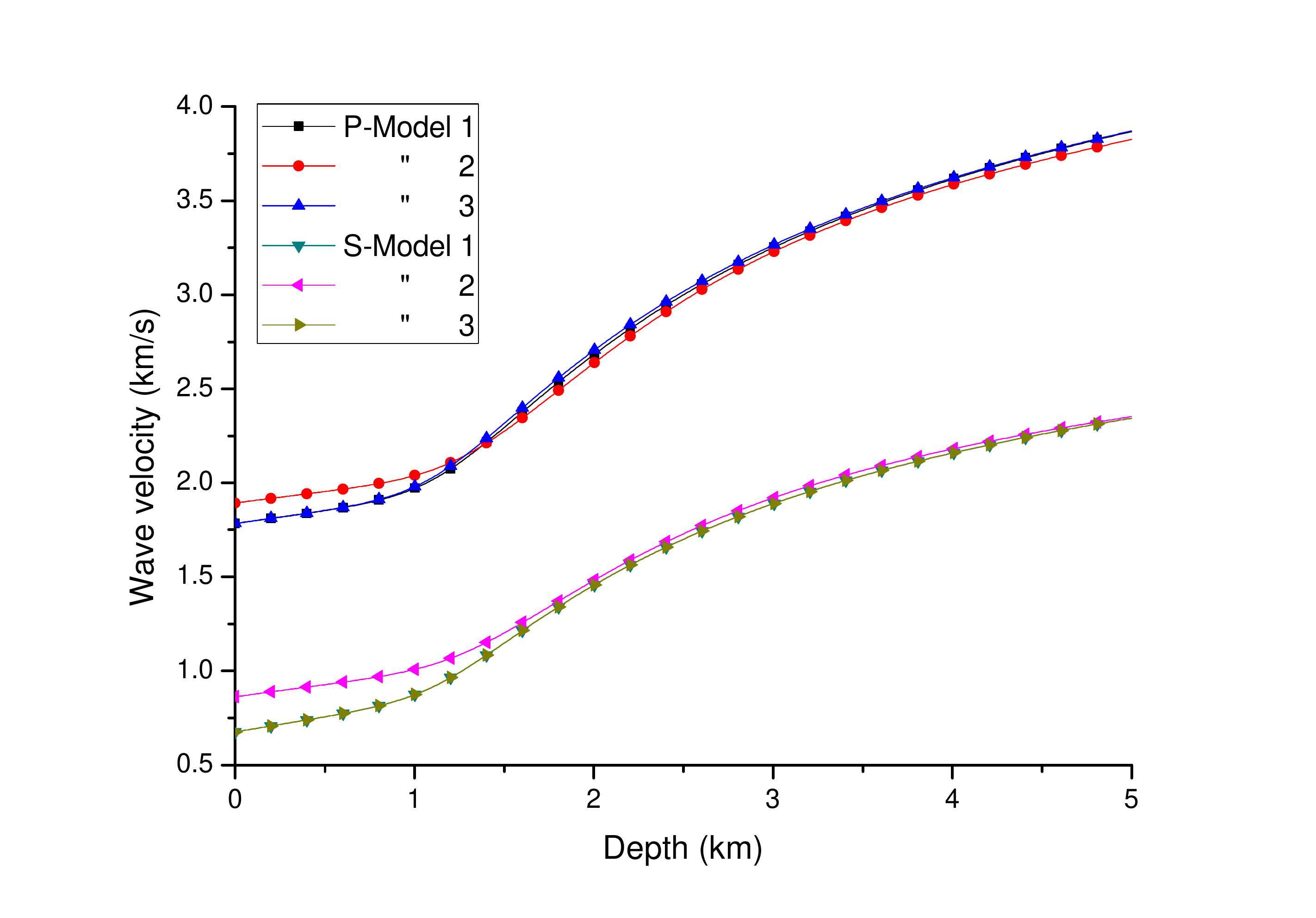}
\caption{
P-wave (a) and S-wave (b) velocities of the shale as a function of depth, corresponding to Models 1, 2 and 3. Model 2 considers a pore aspect ratio $\gamma$ = 0.17.}
\end{figure}

\begin{figure}
\includegraphics[width=11cm]{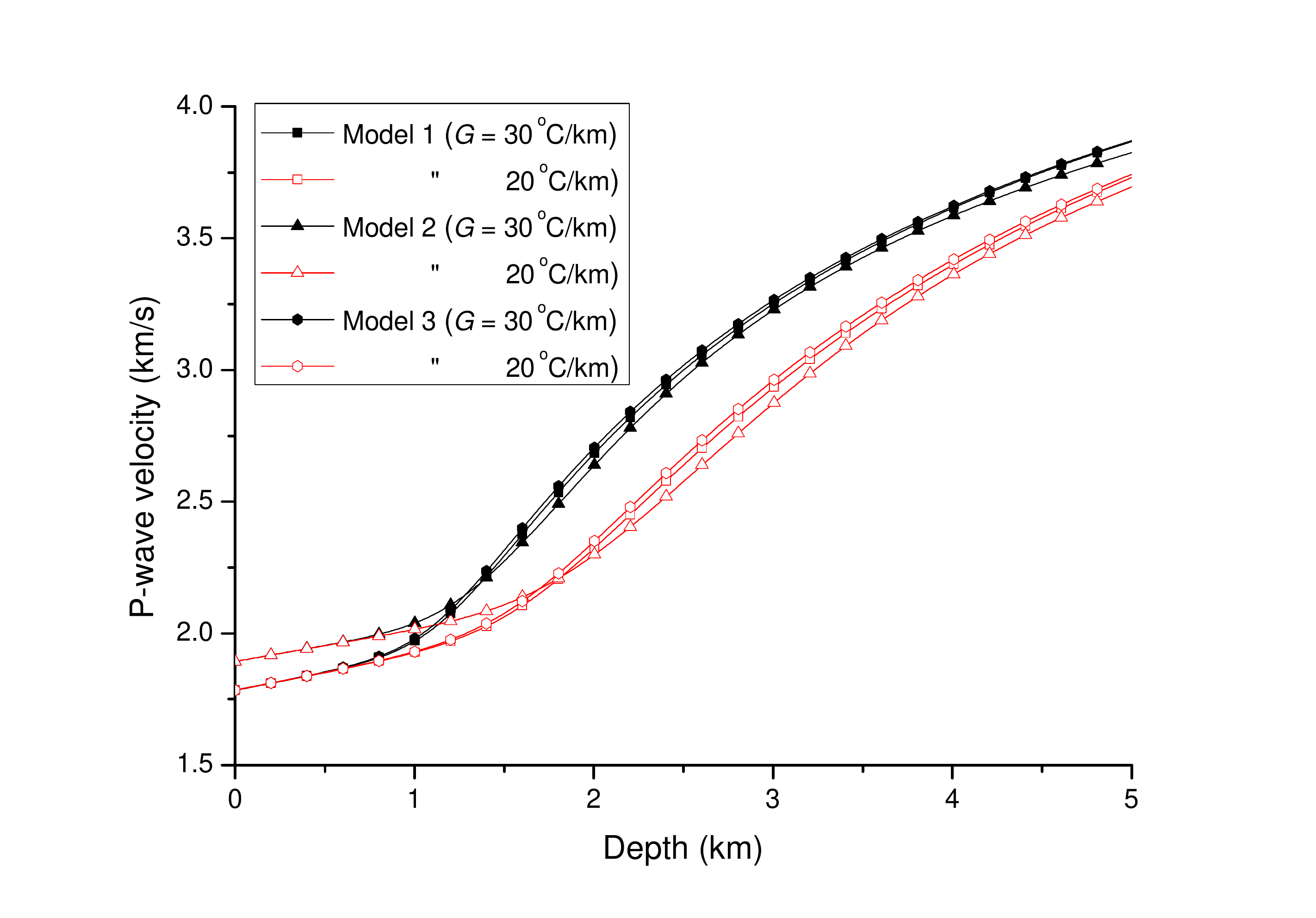}
\caption{
P-wave velocities of the shale as a function of depth, corresponding to Models 1, 2 and 3 for two values of the geothermal gradient.}
\end{figure}

\begin{figure}
\includegraphics[width=10cm]{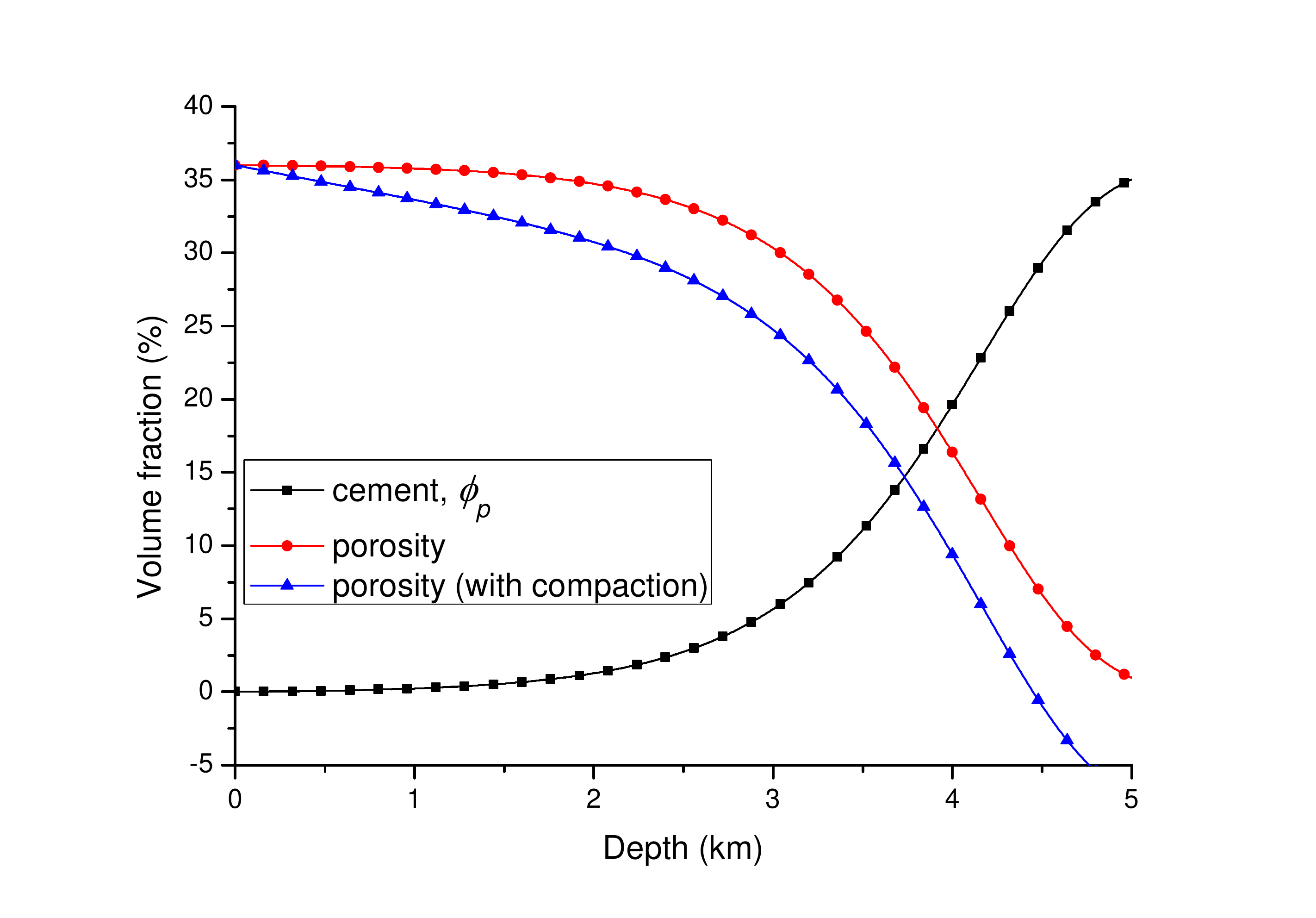}
\caption{
Cement fraction $\phi_p$, porosity (no compaction) and porosity (with compaction) as a function of depth of a sandstone body subject to precipitation and cementation.}
\end{figure}

\begin{figure}
\includegraphics[width=10cm]{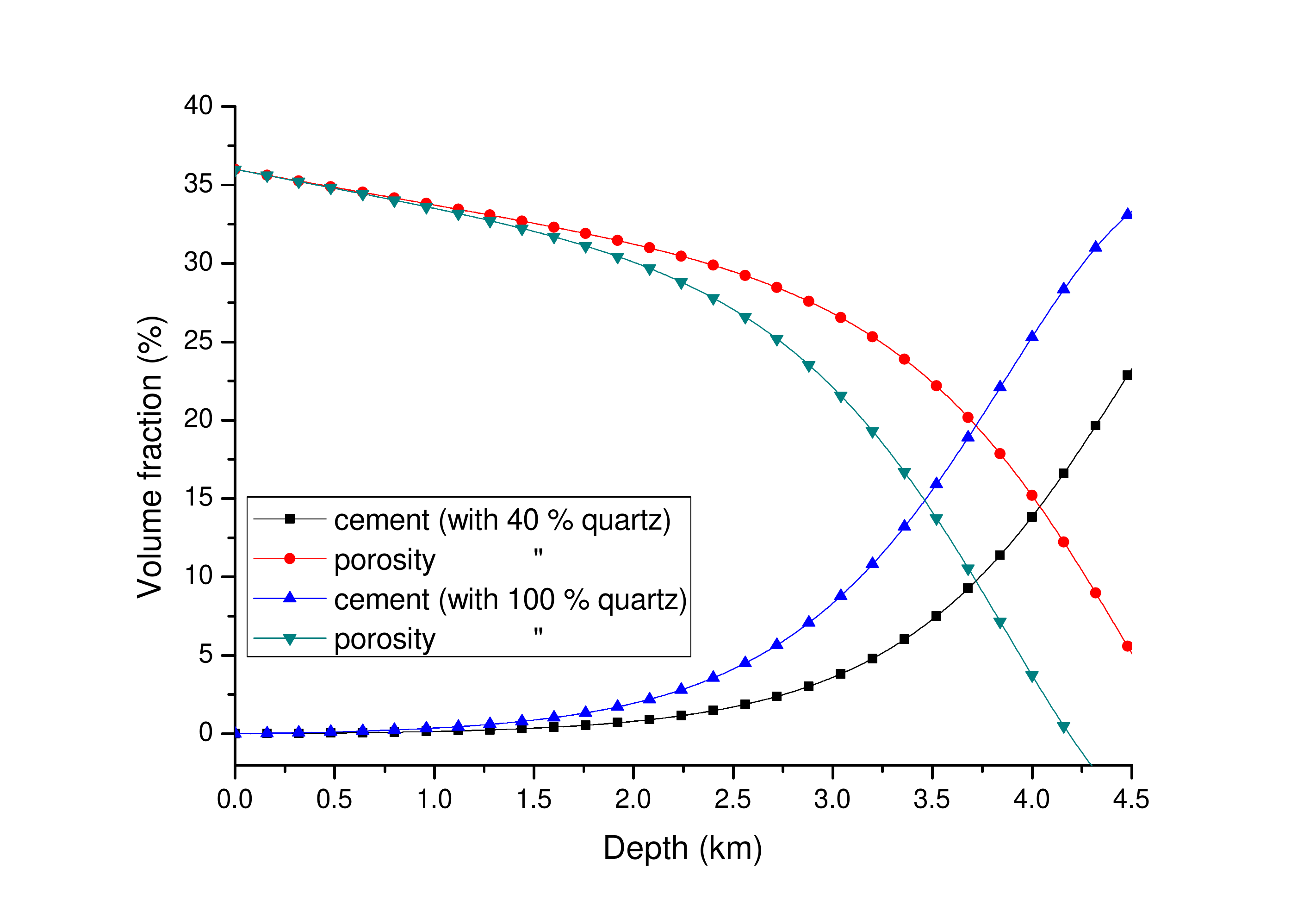}
\caption{
Cement fraction $\phi_p$ and porosity $\phi$ as a function of depth for two values of detrital quart in the sandstone $f$ = 0.4 and $f$ = 1.}
\end{figure}

\begin{figure}
\includegraphics[width=10cm]{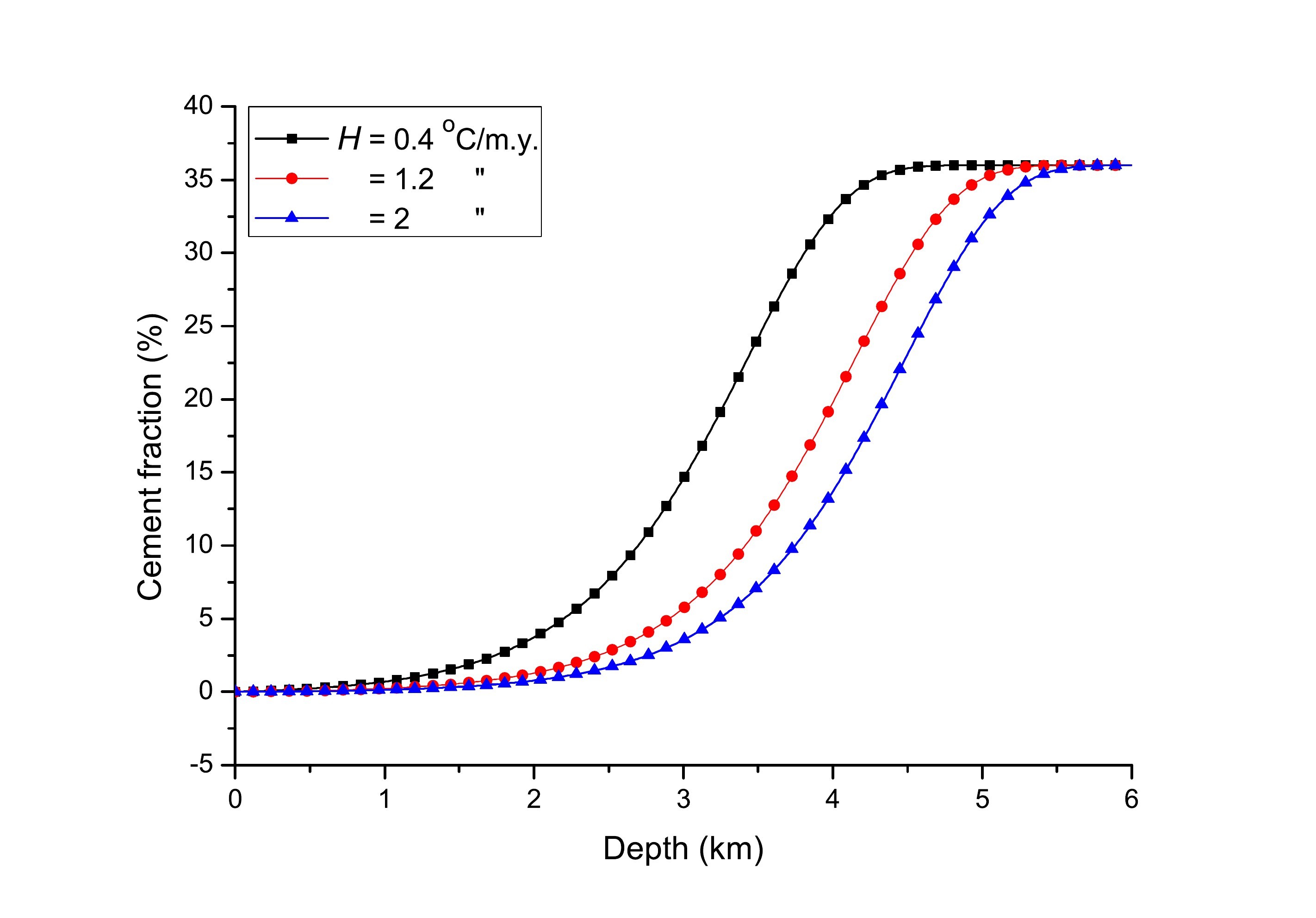}
\caption{
Cement fraction $\phi_p$ as a function of depth for three values of the heating rate.}
\end{figure}

\begin{figure}
\includegraphics[width=10cm]{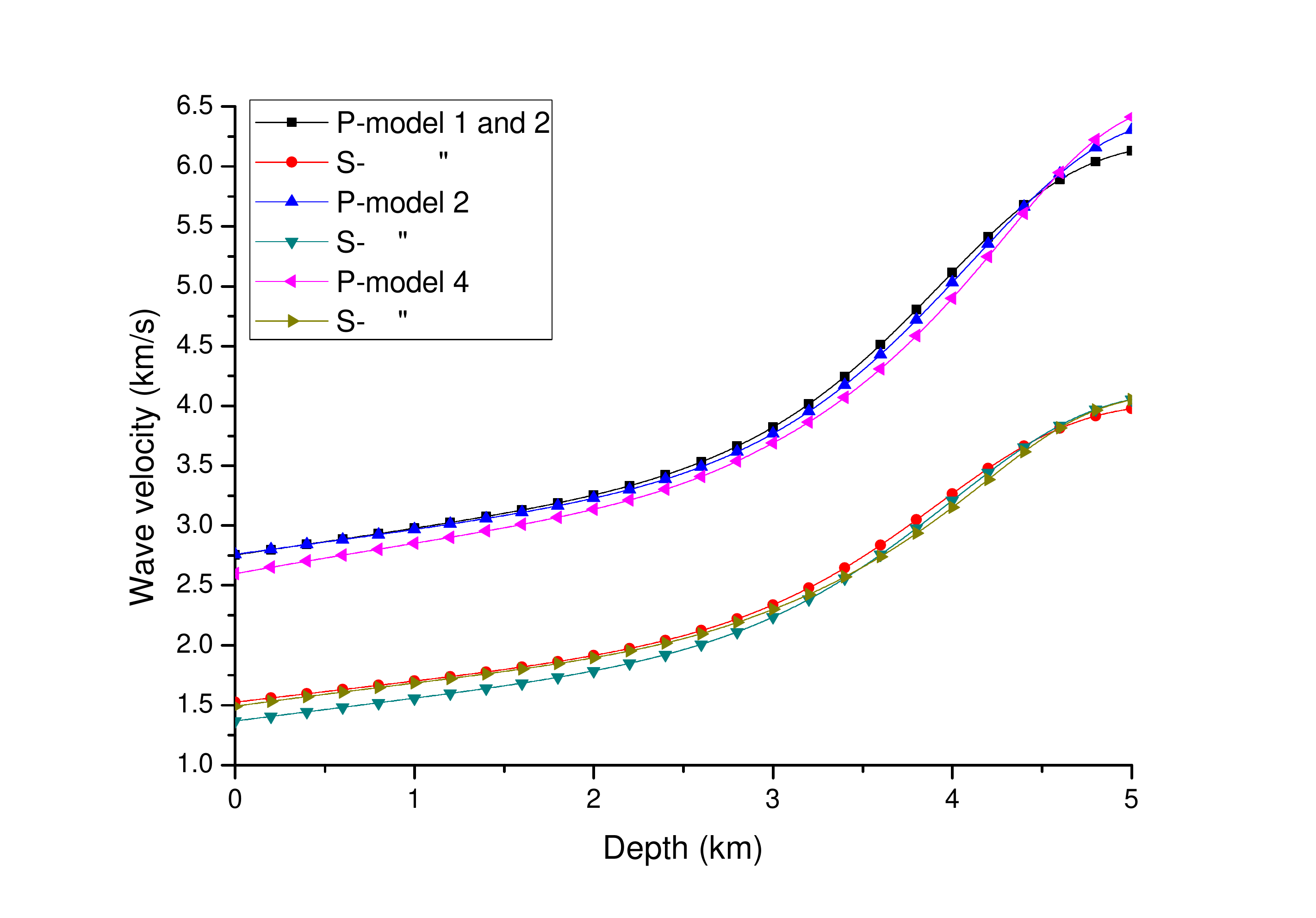}
\caption{
P- and S-wave velocities of the sandstone as a function of depth, corresponding to the four petro-elastical models.}
\end{figure}

\begin{figure}
\hspace{-3cm} \includegraphics[width=15cm]{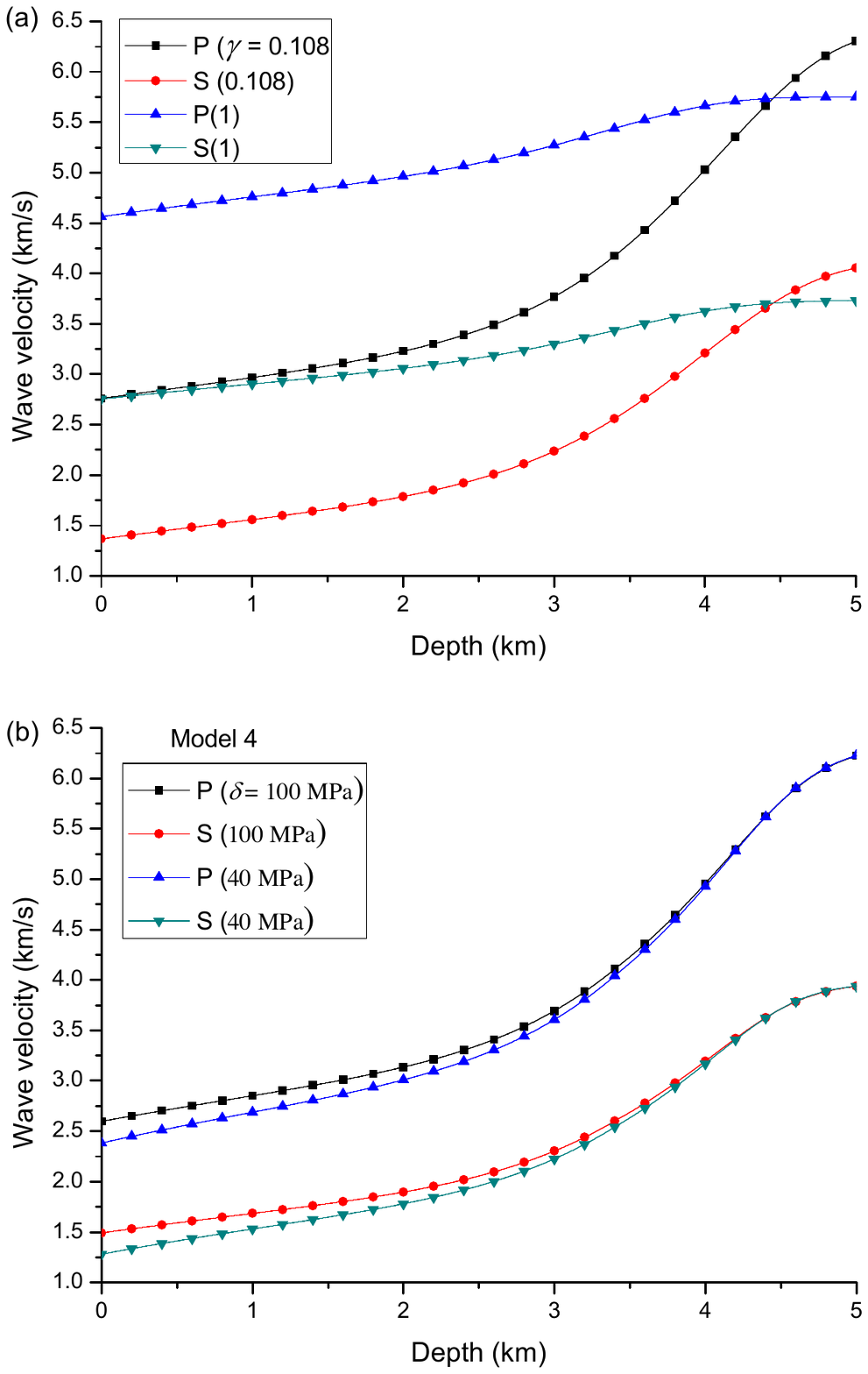}
\vspace{-6cm}
\caption{
P- and S-wave velocities of the sandstone as a function of depth, corresponding to Model 2 (a) and Model 4 (b. for two values of the pore-aspect ratio $\gamma$ and bonding parameter $\delta$, respectively.}
\end{figure}

\begin{figure}
\includegraphics[width=11cm]{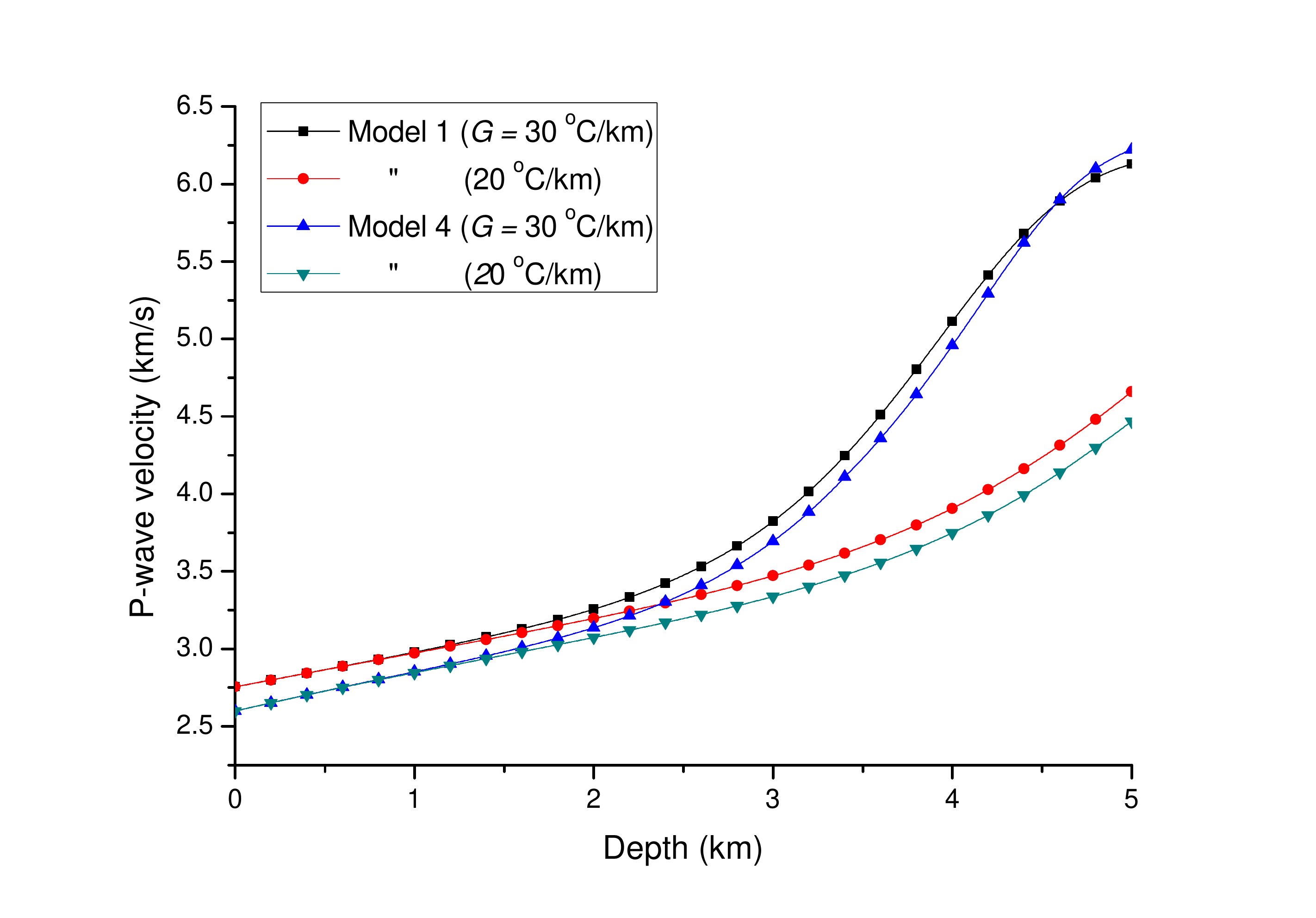}
\caption{
P-wave velocities of the sandstone as a function of depth, corresponding to Models 1 and 4 for two values of the geothermal gradient.}
\end{figure}

\begin{figure}
\includegraphics[width=11cm]{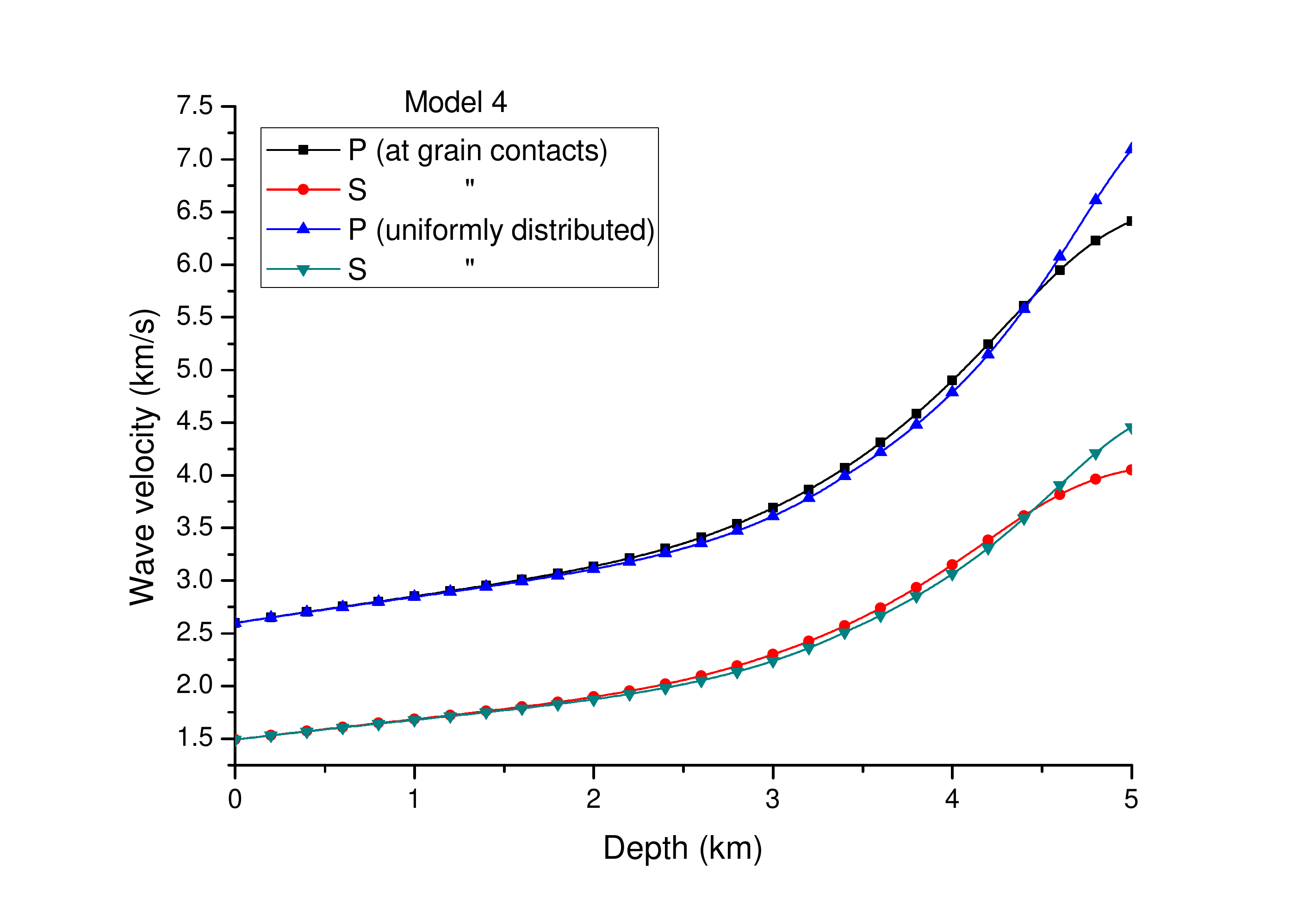}
\caption{
P- and S-wave velocities of the sandstone as a function of depth, corresponding to Model 4 and 
two different distributions of the contact cement, based on equations (\ref{dv2}) and (\ref{dv21}) (cement at grain contacts and coating the grains, respectively).}
\end{figure}

\end{document}